\documentclass[manuscript,screen,hidelinks]{acmart}
\AtBeginDocument{%
  }

\setcopyright{none}
\copyrightyear{2018}
\acmYear{2018}
\acmDOI{XXXXXXX.XXXXXXX}

\acmJournal{TOSEM}

\usepackage{booktabs} %
\usepackage{colortbl}
\usepackage{adjustbox}

\usepackage[strict]{changepage}
\usepackage{xcolor}

\definecolor{darkblue}{rgb}{0.0, 0.0, 0.5}
\usepackage{framed}
\usepackage[most]{tcolorbox}
\tcbuselibrary{skins}
\usepackage{pgf}
\usepackage{collcell}

\definecolor{formalshade}{rgb}{0.95,0.95,1}

\usepackage{xspace}
\usepackage{comment}
\usepackage{multirow}
\usepackage{subcaption}

\newcommand{\rqone}{How performance time series differ from time series from other domains?}

\newcommand{\rqtwo}{How consistent are humans in identifying potential performance changes?}

\newcommand{\rqthree}{How efficient are CPD methods in identifying potential performance changes?}

\newcommand{\ie}{\emph{i.e.,}\xspace}
\newcommand{\eg}{\emph{e.g.,}\xspace}

\newcommand{\etal}{\emph{et~al.}\xspace}
\newcommand{\secref}[1]{Section~\ref{#1}\xspace}
\newcommand{\figref}[1]{Figure~\ref{#1}\xspace}

\newcommand{\tabref}[1]{Table~\ref{#1}\xspace}

\newcommand{\rev}[2]{#2}

\newcommand{\father}{van den Burg and Williams~\cite{Gerrit:2020:ChangePointDetection}\xspace}
\newcommand{\graal}{GraalVM\xspace}
\newcommand{\mongo}{MongoDB\xspace}
\newcommand{\sap}{SAP HANA\xspace}
\newcommand{\potentialchangepoints}{suspected change points~}
\newcommand{\suspiciouschangepoints}{suspected change points~}

\let\oldcite\cite

\renewcommand{\cite}[1]{%
  \ifx\relax#1\relax%
    \textcolor{red}{$\langle$empty citation$\rangle$}%
  \else%
    \if\relax\detokenize{#1}\relax%
      \textcolor{red}{$\langle$empty citation$\rangle$}%
    \else%
      \oldcite{#1}%
    \fi%
  \fi%
}
\begin{document}

\title{Evaluating Change Point Detection Methods for Software Performance Regression Analysis}

\author{{Diego Elias} Costa}
\email{diego.costa@concordia.ca}
\orcid{0000-0001-7084-2594}
\affiliation{%
  \institution{REALISE Lab, Concordia University}
  \city{Montreal}
  \country{Canada}
}

\author{Michele Tucci}
\email{michele.tucci@univaq.it}
\orcid{0000-0002-0329-1101}
\affiliation{%
  \institution{SPENCER Lab, University of L'Aquila}
  \city{L'Aquila}
  \country{Italy}
}

\author{Luca Traini}
\email{luca.traini@univaq.it}
\orcid{0000-0003-3676-0645}
\affiliation{%
  \institution{SPENCER Lab, University of L'Aquila}
  \city{L'Aquila}
  \country{Italy}
}

\author{Daniele {Di Pompeo}}
\email{daniele.dipompeo@univaq.it}
\orcid{0000-0003-2041-7375}
\affiliation{%
  \institution{SPENCER Lab, University of L'Aquila}
  \city{L'Aquila}
  \country{Italy}
}

\author{Thomas Bach}
\email{thomas.bach03@sap.com}
\orcid{0000-0002-9993-2814}
\affiliation{%
\institution{SAP}
\city{Walldorf}
\state{}
\country{Germany}
}

\author{Francois Farquet}
\email{francois.farquet@oracle.com}
\orcid{0000-0002-1337-8024}
\affiliation{%
\institution{Oracle}
\city{Zurich}
\state{Zurich}
\country{Switzerland}
}

\author{David Daly}
\email{ddaly@computer.org}
\orcid{0000-0001-9678-3721}
\affiliation{%
\institution{Independent Researcher}
\country{United States}
}

\author{Simon Eismann}
\email{simon.eismann@uni-wuerzburg.de}
\orcid{0000-0003-1224-4945}
\affiliation{
    \institution{University of Würzburg}
    \city{Würzburg}
    \country{Germany}
}

\author{Petr Tůma}
\email{petr.tuma@d3s.mff.cuni.cz}
\orcid{0000-0002-7035-2322}
\affiliation{%
 \institution{Charles University}
 \city{Prague}
 \country{Czech Republic}
}

\author{Vittorio Cortellessa}
\email{vittorio.cortellessa@univaq.it}
\orcid{0000-0002-4507-464X}
\affiliation{%
 \institution{SPENCER Lab, University of L'Aquila}
 \city{L'Aquila}
 \country{Italy}
}

\author{Andr\'e {van Hoorn}}
\email{andre.van.hoorn@uni-hamburg.de}
\orcid{0000-0003-2567-6077}
\affiliation{%
 \institution{University of Hamburg}
 \city{Hamburg}
 \country{Germany}
}

\renewcommand{\shortauthors}{Costa et al.}

\begin{abstract}
Performance issues in software systems are a critical quality issue that can erode user trust, violate service-level agreements, and ultimately affect business efficiency. 
Consequently, software performance engineering has shifted its focus to developing robust techniques to detect performance regressions as early as possible in the development cycle. 
Performance regression analysis often relies on time series of performance measurements to detect significant changes in performance behavior. 
Change Point Detection (CPD) methods have been widely used to automate the identification of such changes in various domains, including finance, healthcare, and performance monitoring.
However, the effectiveness of these methods for software performance measurements has not been thoroughly evaluated. 
In this paper, we present a comprehensive study to evaluate the effectiveness of various CPD methods on real-world software performance measurement datasets. We start by collecting performance measurement data from three large software systems and characterizing the unique properties of performance time series data. Then, we undertake a large-scale effort to annotate and evaluate the consistency of human annotators' identification of potential performance changes. Thereafter, we evaluate the accuracy of twelve distinct CPD methods in detecting potential performance changes, providing insights into their applicability and effectiveness in software performance regression analysis.
\end{abstract}


\received{20 February 2007}
\received[revised]{12 March 2009}
\received[accepted]{5 June 2009}

\maketitle

\section{Introduction}
\label{sec:introduction}
Performance is a crucial non-functional property of software systems. Performance regressions can directly affect user satisfaction \cite{brutlag2009speed} and increase operational costs, potentially leading to significant economic losses for companies \cite{fastcompany2012amazon}.
To prevent performance regressions from creeping into production, software teams typically rely on continuous performance testing and monitoring~\cite{vokolos1998performance,Jain1991TheAO}. 
This practice involves running the system under a representative workload \cite{Avritzer2002Software} in a pre-production environment and collecting performance-related metrics, such as response times and memory usage.
Given its importance, investment in performance testing infrastructure and automation has steadily increased in recent years~\cite{datainsights2025performance}. Nevertheless, performance testing continues to face several fundamental challenges \cite{Traini2022, DOSSANTOS2025107641,Costa:19:JMH}.
For instance, a crucial factor that hinders the effective application of performance testing in practice is the limited capability to reliably and automatically determine the presence of performance regressions \cite{Fagerstrom2016,Traini2022,Chen2017Analytics, Daly:20:ChangePointDetection}.
Unlike functional tests, which return a clear pass–fail verdict, performance testing results are typically more subtle and difficult to interpret. First, performance measurements are often subject to large variability due to the inherent non-deterministic behavior of software performance \cite{Maricq2018Taming, Papadopoulos2021Methodological, Traini2022Towards}. Second, performance testing outcomes are highly susceptible to the execution environment, hence even slight changes in the testing environment can easily alter the results \cite{Daly:20:ChangePointDetection}.

In this context, merely employing a threshold-based approach---\eg checking whether the mean response time of a performance test on a new software version is 10\% higher than that of the previous version---can easily lead to misleading conclusions. Prior work has shown that such an approach can result in a high number of false positives \cite{Daly:20:ChangePointDetection, Georges2007Statistically}, increasing the human effort required to manually verify whether the reported performance issues warrant further investigation \cite{Daly:20:ChangePointDetection}. This problem is further exacerbated by the scale of today's performance testing infrastructures, which often employ large numbers of tests of various types (\eg microbenchmarks~\cite{Laaber2018evaluation} and load tests~\cite{loadtestingpaper}), executed under different configurations and hardware environments \cite{Traini2022,Daly:21:VirtuousCycle,Rehmann2016Performance}.

\begin{figure*}
    \begin{subfigure}{0.32\textwidth}
        \includegraphics[width=\textwidth]{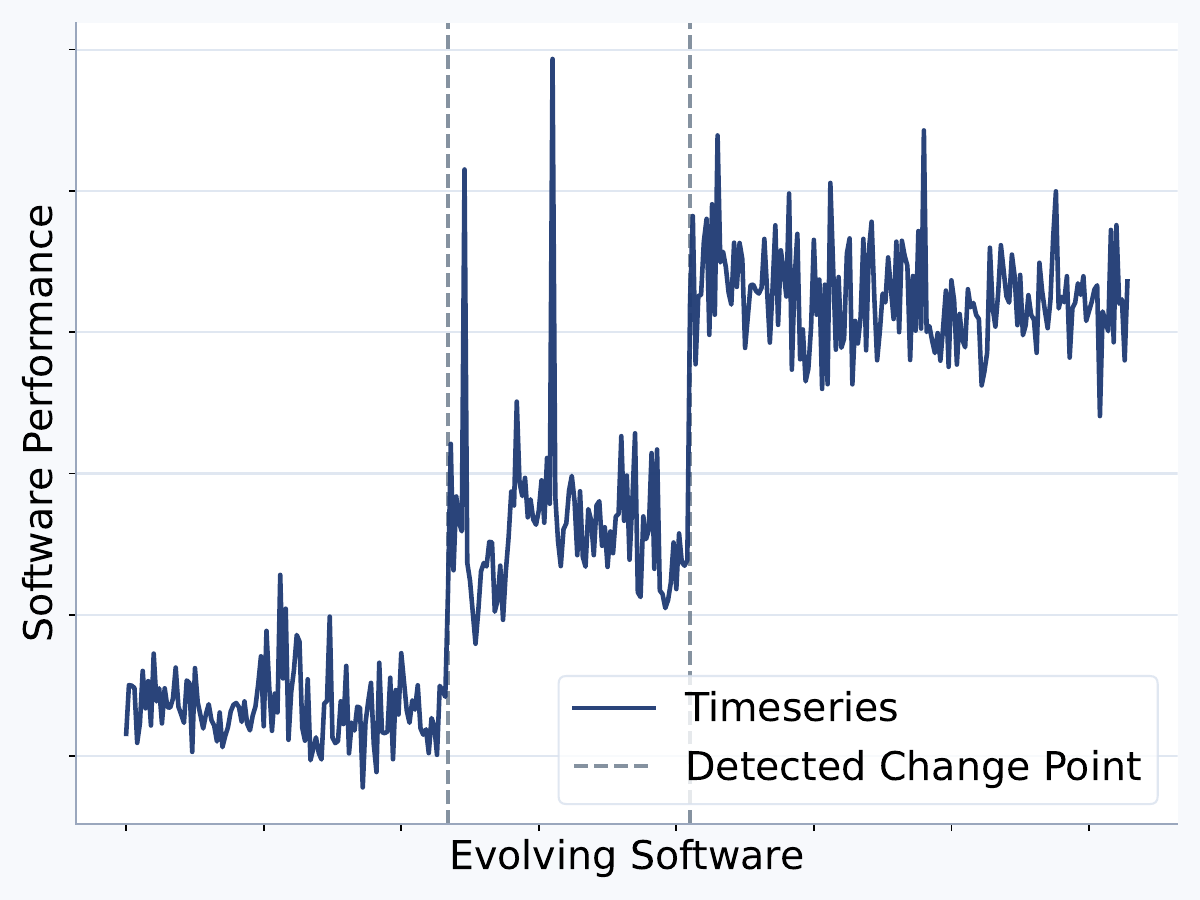}
        \caption{MongoDB}
        \label{fig:exmongo}
    \end{subfigure}
    \begin{subfigure}{0.32\textwidth}
        \includegraphics[width=\textwidth]{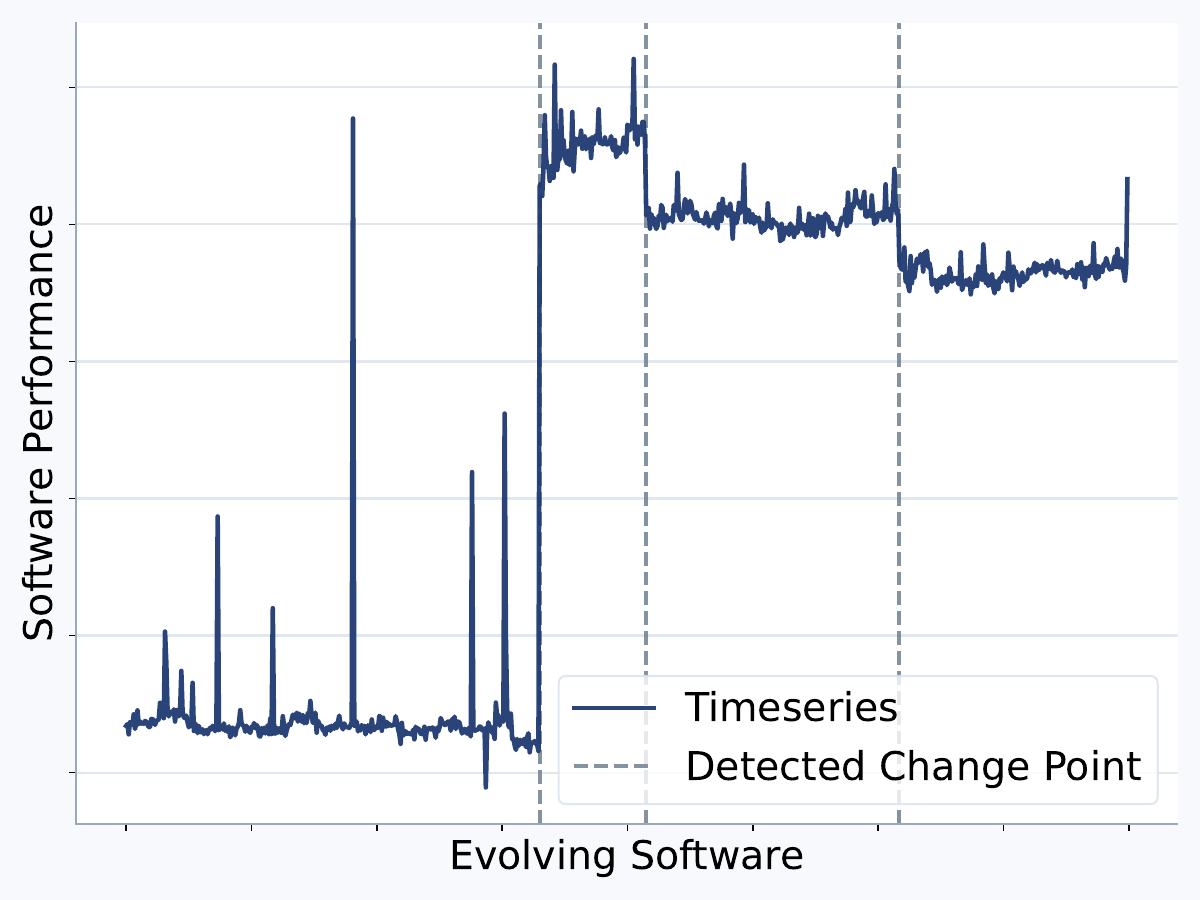}
        \caption{SAP}
        \label{fig:exsap}
    \end{subfigure}
    \begin{subfigure}{0.32\textwidth}
        \includegraphics[width=\textwidth]{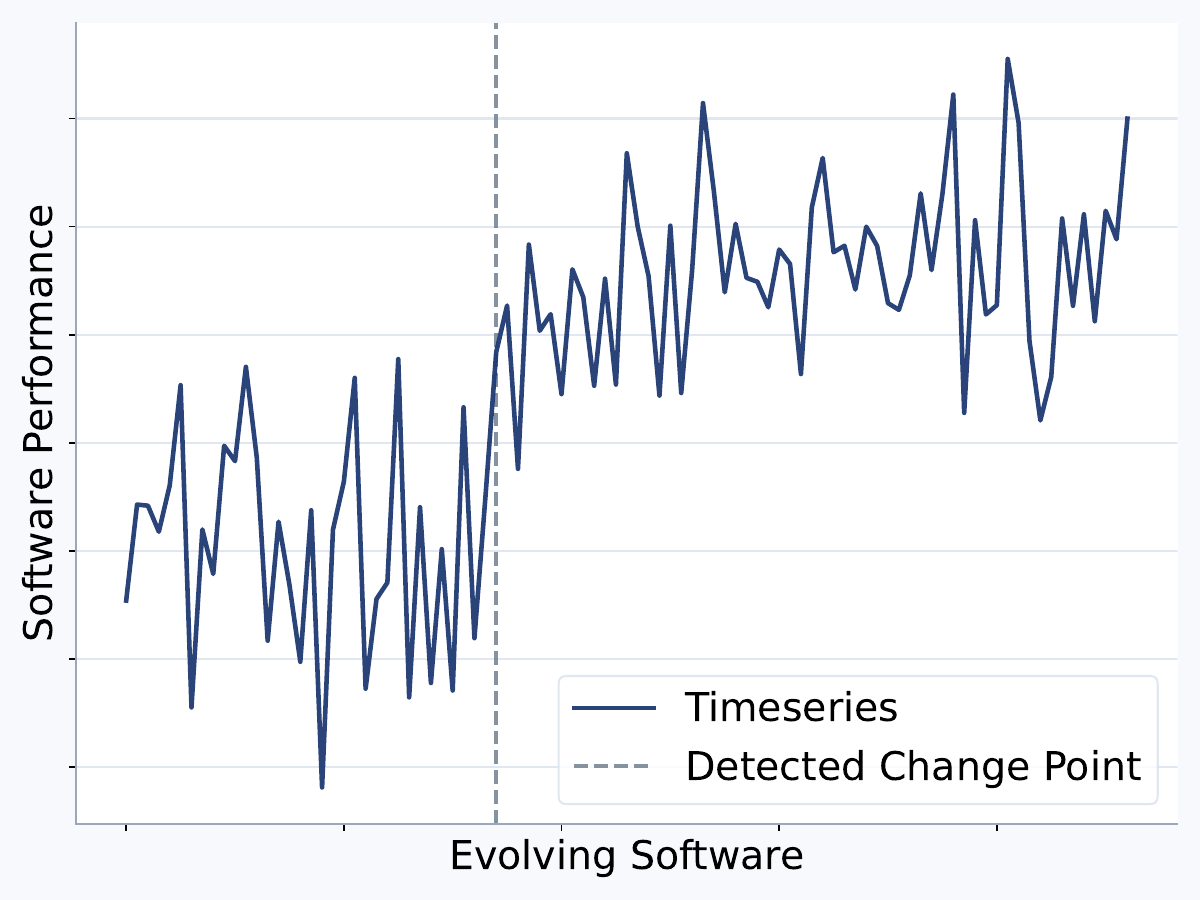}
        \caption{GraalVM}
        \label{fig:exgraal}
    \end{subfigure}  
    \caption{Examples of performance time series from different performance testing infrastructures. Vertical dotted lines denote change points identified by the E-Divisive Means algorithm~\cite{Matteson2014, Daly:20:ChangePointDetection}.%
    }
    \label{fig:extimeseries}
\end{figure*}

To address this challenge, practitioners have begun employing change point detection (CPD) methods~\cite{Aminikhanghahi2017survey} to detect suspicious performance regressions with greater statistical rigor~\cite{Daly:20:ChangePointDetection,Fleming2023Hunter}. 
In these approaches, the results of performance tests across multiple software versions are ordered sequentially as a time series, where each data point represents the outcome of a performance test (\eg mean response time) for a specific software version. %
One of the first pieces of evidence supporting this methodology was presented by Daly \etal \cite{Daly:20:ChangePointDetection}, who proposed using the E-Divisive Means algorithm~\cite{Matteson2014}---a well-known CPD algorithm---to identify abrupt changes in a time series of MongoDB performance test results. According to their findings, this methodology dramatically reduced the false positive rate in detecting performance changes, while also making the overall performance evaluation process more manageable. Subsequently, similar approaches have also been deployed in other companies \cite{Fleming2023Hunter}. Figure \ref{fig:extimeseries} shows a few examples of the application of such approaches.

Despite promising results, the application of CPD for performance change detection remains fundamentally restricted to one specific CPD algorithm, even though the literature provides a wide range of statistical CPD methods \cite{Truong2020} that could potentially be applied to this task. At present, there remains limited knowledge about the effectiveness of other CPD algorithms for performance change detection. Furthermore, the existing evidence on the effectiveness of these techniques is limited to a few industrial experience reports, and no rigorous or extensive empirical evaluations have yet been conducted. This paper aims to fill these gaps.

To achieve our research objective, we first gathered time series of performance measurements, hereby referred by \textit{performance time series}, 
from the performance testing infrastructures of three large software companies: Oracle, SAP, and MongoDB. 
We explored the characteristics of performance time series and compared them against time series from four other domains to shed light on the differences and similarities. 
We then developed a rigorous annotation process involving 14 performance engineering experts, who manually identified suspicious performance changes on a sample of 1,500 time series. Finally, we used the resulting annotations to evaluate the effectiveness of 12 CPD algorithms in detecting potential performance changes.

Results show that (i) performance time series show distinctive statistical properties when compared to time series from other relevant domains; (ii) human annotators show a high level of agreement in identifying suspected performance changes, although they remain susceptible to some degree of subjectivity; and (iii) the effectiveness of CPD methods varies considerably depending on the specific performance testing infrastructure, with $F_1$-score improving substantially--by 28\% on average--when the methods are optimally tuned.

The contributions of this work are fourfold:
\begin{enumerate}

    \item We characterize software performance time series and compare their statistical properties with those of time series from other relevant domains (e.g., nature and finance).

    \item We conduct extensive manual labelling work to establish a large dataset of real systems' suspected performance changes. 
    This dataset can encourage further development of CPD  methods tailored for performance measurement data, and it is made publicly available, along with the scripts used for the analyses in this work\footnote{\url{https://doi.org/10.5281/zenodo.18889836}}. 

    \item We have adapted the time series labelling tool proposed by~\father to performance time series. The new version includes support for difficulty evaluation, a tailored tutorial for performance time series, and new annotation types. 
    
    \item We provide the most extensive evaluation to date of CPD methods for identifying performance regressions in software systems, involving 12 different algorithms and three real-world datasets of performance time series.
\end{enumerate}

\section{Motivation}
\label{sec:motivation}
The performance of a large software system is continuously tested to ensure that it remains responsive to its users, cost-effective, and compliant with the defined service-level agreements (SLA). 
There are various performance testing practices established in the industry, such as testing the system under expected user load (load testing) or evaluating the system's robustness by testing it beyond normal operations (stress testing)~\cite{DBLP:conf/icsm/MeglioSPORM25,DBLP:journals/tse/WeyukerV00}. 
In all such cases, performance measurements are aggregated as time series, hereafter referred to as \textit{performance time series}, where practitioners can analyze the evolution of their system's performance during the history. 
However, analyzing such performance time series is a non-trivial task, as it requires specialized techniques and algorithms to identify performance flaws.
To the best of our knowledge, little even less comprehensive research has been conducted on analyzing performance time series to identify performance flaws~\cite{Daly:20:ChangePointDetection,fleming2023hunterusingchangepoint}.
Therefore, in this paper, we describe the fundamental concepts of change point detection methods and then address some key questions related to analyzing performance time series. 

\subsection{Change Point Detection methods}

Change point detection (CPD) methods aim at identifying moments of abrupt change in the behavior of a time series. 
CPD is a crucial activity of data analysis, and it has many practical applications in finance~\cite{Lavielle2007}, business~\cite{Taylor2018}, quality control~\cite{Page1954}, network traffic analysis~\cite{Kurt2018}, and many other domains.
In the past decades, several automated algorithms have been developed to accurately determine \emph{change points} in time series data (see \cite{Truong2020} for a detailed review).
These algorithms partition the input time series into a number of segments, where each segment identifies a clear shift in the behavior of the series.

Given a time series $y_1, y_2, ..., y_T$, where $y_t$ denotes the observation for time step $t$ and $T$ the length of the series, a CPD algorithm returns an ordered set of time steps $\mathcal{T} =\{\tau_0,\tau_1, ..., \tau_n \}$,  where each change point $\tau_i$ marks the first observation of a new segment in the partition. For notational convenience, it is usual to define $\tau_0=1$ and $\tau_{n+1} = T + 1$.
In CPD, the optimization problem is typically given by the following objective function:

\begin{equation}
	\min_{\mathcal{T}}{\sum_{i=0}^{\tau_{n+1}}\ell(y_{\tau_{i-1}:\tau_{i}-1}) + \lambda P(n)} 
\end{equation}
where $y_{\tau_{i-1}:\tau_{i}-1}$ denotes a segment of the series starting at time step $\tau_{i-1}$ and ending at time step $\tau_{i}-1$, $\ell(\cdot)$ denotes a cost function for a segment (e.g., the negative maximum log-likelihood), $\lambda\geq 0$ a hyperparameter, and $P(n)$ a penalty on the number of change points~\cite{Killick2012}.

\section{Related Work}
\label{sec:related}

In this section, we review the literature related to two key questions regarding the analysis of performance time series: the consistency of human experts in flagging suspicious performance changes, and the efficiency of CPD methods in identifying such changes.

\subsection{How consistent are experts in flagging suspected performance changes?}

Performance regression detection often relies on human experts to analyze performance metrics and identify suspected changes. 
The human-in-the-loop approach leverages domain knowledge and experience to interpret complex performance data, and experts' knowledge could bias the analysis.
In addition, as software systems grow in complexity and size, the volume of performance data increases, making manual analysis more challenging (even overwhelming) and error-prone~\cite{foo2010mining,miningsoftrepo}.
\citet{foo2010mining} reported that experts often show inconsistency when flagging suspicious performance changes due to subjective judgment and lack of formal baselines.
\citet{miningsoftrepo} showed that manual analysis of metrics measurements from performance tests is often inconsistent, as analysts rely on subjective judgment in the absence of commonly recognized criteria. 
This leads to variability in detecting regressions and risks both false positives and missed issues.
They also emphasized that automated methods reduce analysis time, human bias, and improve consistency.

Consistency and reliability of expert evaluations in identifying performance regressions heavily rely on the context and complexity of the system under test, the quality of the performance data, and the expertise of the evaluators themselves~\cite{fleming2023hunterusingchangepoint,Besbes_2025,perfabnormalities,loadtestingpaper}.
\citet{Besbes_2025} stated that once generated by an automated detection system, performance anomalies must carefully be investigated and validated by performance experts who should link these anomalies to corresponding bugs for fixing.
\citet{perfabnormalities} exploited visuals graphs for performance measurements to identify performance abnormalities.
They highlighted the needs for human expertise to interpret complex graphs and identify subtle performance issues that automated systems might miss.
\citet{loadtestingpaper} found that manual detection of performance regressions, specifically in the case of load testing, is error-prone and inefficient, as analysts must compare counters to thresholds and shift through logs under time pressure and limited system knowledge.

The literature indicates that although experts might be inconsistent in flagging suspicious performance changes, their expertise remains crucial for interpreting complex performance data and validating automated detection results.
Therefore, we asked human experts to label performance time series from real-world systems to create a ground truth dataset for evaluating CPD methods.

\subsection{How efficient are CPD methods in identifying suspected performance changes?}

Performance measurements are usually gathered over time to monitor and evaluate the performance of software systems.
Thus, representing them as time series allows practitioners to analyze the evolution of such performance metrics across different software versions.
Change point detection (CPD) methods can be employed to identify significant shifts in these time series, which may indicate potential performance regressions or improvements.
Therefore, it is of vital importance to evaluate the efficiency of CPD methods in identifying suspicious performance changes in real-world scenarios.

\citet{reichelt2018} compared several statistical tests, such as T-test~\cite{boneau1960effects}, G-test~\cite{woolf1957log}, and Mann-Whitney U-test~\cite{mann1947test} for detecting performance changes. 
From  \citeauthor{reichelt2018} results emerged that T-test maintained over 99\% precision with over 80\% recall, making it the most balanced and efficient choice.
\citet{muehlbauer2019accurate} and \citet{lee2012} explored the use of Cumulative Sum (CUSUM)~\cite{koshti2011cumulative} to identify performance anomalies in software systems.
\citet{muehlbauer2019accurate} identified performance evolution of six real-world software systems, and \citet{lee2012} focused on database systems during continuous integration.
\citeauthor{muehlbauer2019accurate} and \citeauthor{lee2012} found that CUSUM could reliably detect small performance changes while maintaining a low false alarm rate.

Along with the statistical methods, (un)supervised ML-based and probabilistic modeling CPD methods have also been explored for performance change detection.
For example, \citet{pcachartcontrol} demonstrated that supervised approaches are generally more stable and effective than unsupervised methods when multiple performance counters are used for system performance analysis. Moreover, supervised techniques are better suited to identifying the most relevant performance counters for CPD. However, this advantage comes at the cost of increased overhead due to their semi-automated training phase.
\citet{garnett2010} analyzed a Bayesian change point detection method by marginalizing change point locations as hyperparameters. Thus, they have proved that the Bayesian method could effectively identify sudden changes in performance time series.
\citet{DBLP:conf/icsm/Lyu0LNH24} combined PELT~\cite{Killick2012} with Bayesian optimization in a hybrid framework aimed at processing large volumes of performance data in modern database systems to minimize false negatives and their associated alarms. 
\citet{cloudplatformapp} showed that Optimal Partitioning (e.g., PELT~\cite{Killick2012}), Heuristic methods (e.g., Binary Segmentation~\cite{chen1993joint}), and Non-parametric algorithms (e.g., E-Divisive mean~\cite{DBLP:conf/sigmod/IngoD20}) can detect sustained workload changes and can help attribute anomalies to root causes.
\citet{fleming2023hunterusingchangepoint} proposed a modified version of E-Divisive mean algorithm that outperformed PELT in detecting performance regression in a synthetic software performance dataset.

\smallskip

From the literature we investigated, it emerges that CPD methods can effectively identify suspicious performance changes in real-world scenarios.
However, the efficiency of these methods can vary based on the specific characteristics of the performance time series. 
Therefore, it is crucial to evaluate and compare different CPD methods on diverse datasets to determine their effectiveness in various contexts.
To address this need, we evaluated a comprehensive set of CPD methods on performance time series from real-world software systems, providing insights into their strengths and limitations in detecting performance changes.

\section{Study Design}

\begin{figure}
    \centering
    \includegraphics[width=1\linewidth]{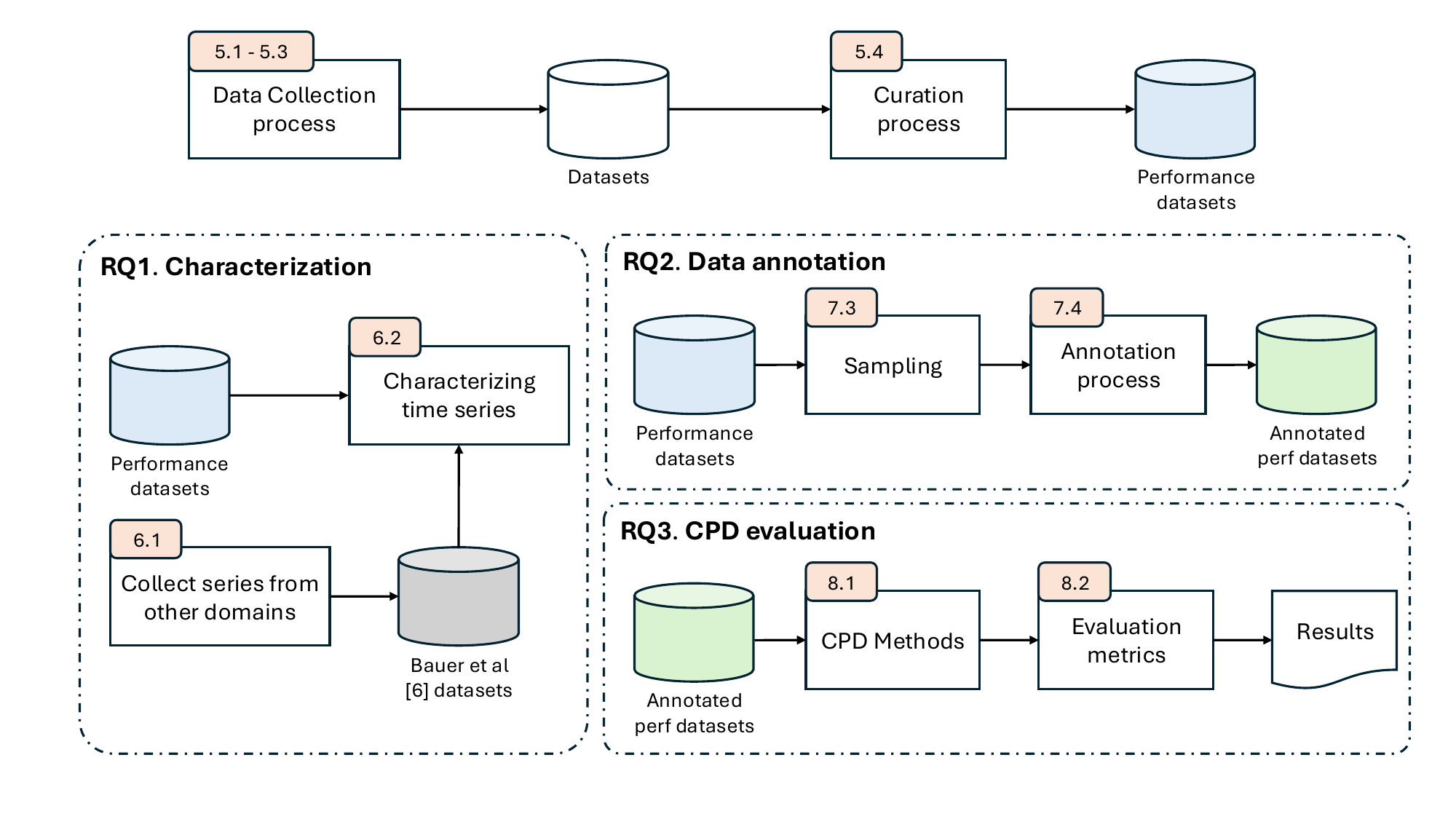}
    \caption{Study design and methodology overview. The annotated number above each process refers to its respective section in the paper.}
    \label{fig:study-design}
\end{figure}

The goal of our study is to compare the effectiveness of well-established Change Point Detection methods in identifying software performance regressions. 
To accomplish our goal, we design a study that explores the characteristics of performance time series, annotates data used for CPD evaluation, and experimentally compares the state-of-the-art CPD methods. 
We show an overview of our study design in Figure~\ref{fig:study-design}.
As shown in our overview, we start by collecting performance data, and in particular, we collect time series from three large-scale industrial systems: MongoDB, GraalVM, and SAP Hanna. 
Using these datasets, we structure our study methodology to answer three main research questions: 

\begin{itemize}
    \item \textbf{RQ1.} \textbf{\rqone}
    As the effectiveness of CPD methods is tied to the characteristics of the time series~\cite{Wang:09:Characteristics}, we start our analysis by characterizing and exploring the differences between the collected performance time series and time series from other domains, such as finance and weather data~\cite{Bauer:Libra:benchmark}. 

    \item \textbf{RQ2.} \textbf{\rqtwo}
    We conduct a large-scale qualitative effort to label potential performance changes. 
    Building on top of the annotation platform proposed by~\father, we select 14 experts in software performance research to annotate 1,500 performance time series and evaluate the convergence and quality of the annotations. 

    \item \textbf{RQ3.} \textbf{\rqthree}
    Using the annotated data and the consensus from RQ2, we conduct an experimental study comparing 12 CPD methods for their effectiveness in identifying potential performance changes. 
\end{itemize}

\section{Datasets}
\label{sec:datasets}

The datasets included in this work contain performance measurements of large software systems. 
In this section, we present an overview of each dataset and the methodology used to establish a reasonably homogeneous dataset for conducting our study.

\subsection{\graal performance measurements dataset}
\label{sub:graal-dataset}

\graal is a high-performance, polyglot virtual machine developed by Oracle that supports multiple programming languages~\cite{graalvm}. 
We collect the performance measurements of the Graal Virtual Machine (GraalVM) using its public API~\cite{GraalVMDataset}.
The public API provides an entry point for collecting \graal performance data through a REST API, with parameters to filter the performance measurement, including the starting date of the performance measurement, so effectively allowing to control the length of the performance time series. 
We collect data for two years, from October 2019 to October 2021.

The \graal dataset contains fine-grained performance measurements. 
Each performance benchmark is run uninterruptedly for 5 to 10 minutes, repeating each run tens to hundreds of times, yielding one data point for each repetition. 
Data points are then typically aggregated to give an overview of the performance measurement of a system component.   
However, data should be carefully aggregated, as micro-benchmarks in Java are notoriously affected by external effects and the warm-up period of the JVM~\cite{Costa:19:JMH}.
To address this issue, we aggregate the results of the \graal benchmark run as follows:

\begin{itemize}
    \item \textbf{Remove the warm-up period.} We approach this problem very conservatively. 
    Guided by the experienced \graal performance engineers, and in relative abundance of measurement data, we consider the entire first half of the measurements to be the warm-up period and remove it from the dataset.
    
    \item \textbf{Aggregate time measurements per system version.} 
    We aggregate the remaining half of the measurements using the average, consolidating them into a single system version measurement. 
    That is, in the \graal dataset, each data point in a timeseries is the result of averaging a single benchmark run.  
    
\end{itemize}

\subsection{MongoDB performance measurements dataset}
\label{sub:mongodb-dataset}

MongoDB runs performance tests in its continuous integration system, with hundreds of tests run under multiple system configurations. 
The performance team at MongoDB uses change point detection methods to identify changes in performance~\cite{Daly:20:ChangePointDetection,Daly:21:VirtuousCycle}, which are then manually triaged and assigned to developers to fix.
The MongoDB performance measurements dataset was made public before our study, as it was part of a Data Challenge track in the International Conference of Performance Engineering in 2022~\cite{MongoDB_dataset}.
Hence, the collection process was simplified by only downloading the Zenodo repository. 
After collecting the data, we performed the following data curation step: 

\begin{itemize}
    
\item \textbf{Focus on the most relevant systems.} 
    The original dataset contains performance measurements of 25 different systems. 
    To keep our analysis more manageable, we focus our study on the most relevant systems for the MongoDB performance team. 
    Following the advice of MongoDB engineers, we selected the measurements for the systems 'sys-perf', 'performance', 'sys-perf-4.4', 'performance-4.4'. 

\end{itemize}

\subsection{SAP performance measurements dataset}
\label{sub:sap-dataset}

SAP HANA is an in-memory database management system targeted for enterprise scenarios~\cite{Bach2022}.
During this study, SAP has made this dataset public at the SAP GitHub repository~\cite{SAP_dataset}.
Hence, we collected the original dataset by downloading the repository. 
The original SAP dataset contains 7522 performance time series with measurements collected during SAP HANA development over a time frame of approximately half a year. 
The dataset contains 84 measurements, from elapsed time and throughput to other system indicators, such as the number of I/O operations performed and the commit rate. 

\begin{itemize}
    \item \textbf{Aggregate time measurements per system version.}  
    Similarly to the \graal dataset, the SAP dataset often contains multiple measurements per system version. 
    We aggregated the measurements related to the same system version by averaging the measurement values.  
\end{itemize}

\subsection{Curation process}\label{sub:curation-process}

The datasets of performance time series included in our study are heterogeneous and may include different types of metrics.  
We opted to further curate our dataset to establish a common ground for comparison, as follows:

\begin{enumerate}
    
    \item \textbf{Focus on the time-based performance metrics.}
    The datasets contain various metrics, including compilation time, memory usage, time spent during garbage collection, and more.  
    As we want to evaluate change point detection methods for identifying performance regression related to system runtime, we filter only tieme series that either include output execution time (time per operation) or throughput (number of operations per time), as both are interchangeable. To homogenize the time series, we converted all throughput values to elapsed time.

    \item \textbf{Excluding short time series.} 
    We want to compare the effectiveness of CPD methods on performance time series with reasonable historical data; hence, we removed any performance time series that contains fewer than 50 performance measurements (data points). 
    The effectiveness of change point detection methods is directly related to data abundance, and methods have been reported to underperform in time series with fewer than 30 data points, due to insufficient sample size for parameter estimation~\cite{Aminikhanghahi2017}. 
    
    \item \textbf{Excluding time series with potential data problems.} 
    We removed from the study any performance time series that contain  measurements with zeroes as values, as we consider a measurement of zero in elapsed time to be an indication of incorrect performance measurement data.
\end{enumerate}

Table~\ref{tab:dataset-overview} shows the outcome of our curation steps applied to all the datasets. 
Our curation process yields 253 time series for the Graal dataset, 14,507 time series for the MongoDB dataset, and 6,554 time series for the SAP dataset.

\begin{table}
    \centering
    \caption{Statistics of the Curated Datasets}
    \label{tab:dataset-overview}
    \begin{tabular}{l r rrr rrr}
\toprule

    \multirow{2}{*}{\textbf{Dataset}}
    & \multirow{2}{*}{\textbf{\# Series}}
    & \multicolumn{3}{c}{\textbf{\# Measurements}}
    & \multicolumn{3}{c}{\textbf{Period in days}} \\
    \cmidrule(lr{0.2em}){3-5}
    \cmidrule(lr){6-8}
 
    & 
    & min
    & avg
    & max
    & min
    & avg
    & max \\

\midrule

    \graal 
    & 506
    & 66
    & 138
    & 193
    & 699
    & 732
    & 819 \\

    \mongo  
    & 14,507
    & 31
    & 221
    & 1422
    & 7
    & 111
    & 787 \\
    
    \sap 
    & 6,581
    & 84
    & 1307
    & 1667
    & 7
    & 180
    & 180 
    \\

\bottomrule
\end{tabular}

\end{table}

\section{Performance Time Series Characteristics}
\label{sec:characterization}

\noindent
Performance measurements are notorious for containing noise and shift patterns that pose challenges to performance regression analysis~\cite{Wang:09:Characteristics,Aminikhanghahi2017}. 
However, the temporal patterns and structural properties of performance time series remain underexplored compared to other domains like finance, and weather forecast, where change point detection is widely applied~\cite{Aminikhanghahi2017survey}.
In this section, we aim to characterize performance time series in contrast to time series from other domains, to identify the applicability and limitations of existing CPD methods.

\subsection{Selecting Time Series from Other Domains}

We start by collecting time series from previous studies, aiming to find a dataset that contains a comparable number of time series collected from different domains. 
While numerous studies have made their dataset available for inspection, to the best of our knowledge, it is rare to find studies that have published cross-domain data.
For those reasons, we opted to select the Libra benchmark dataset, proposed by Bauer \textit{et al.}~\cite{Bauer:Libra:benchmark}.
The Libra benchmark is a time series forecasting benchmark designed to contain a variety of time series collected from different domains and to establish a plain-level comparison across methods. 
The dataset contains 400 heterogeneous time series, from 50 different sources, across four large domains:

\begin{itemize}
    \item \textbf{Economics}: This domain includes macro- and micro-economic indicators such as sales volumes, gas consumption, and unemployment rates. These time series typically exhibit pronounced trends, business-cycle effects, and seasonal components.
    \item \textbf{Finance}: The finance domain comprises time series related to stock prices, sale prices, and exchange rates. Compared to economic indicators, these series often display higher volatility, weaker seasonality, and abrupt regime changes.
    \item \textbf{Human Access}: This domain captures measurements related to human activity and digital infrastructure usage, including Internet traffic, SMS exchanges, and phone call volumes. Such time series are characterized by high frequencies, strong daily or weekly seasonality, and sudden spikes.
    \item \textbf{Nature and demographics}: This domain includes time series associated with environmental and population-related phenomena, such as rainfall levels, birth rates, and death rates. These series often combine long-term trends with seasonal or cyclical patterns and may exhibit domain-specific irregularities.
\end{itemize}

As each domain may exhibit distinct characteristics, we report on the differences between all considered time series, namely our three performance time series datasets and the four domains collected from the Libra dataset~\cite{Bauer:Libra:benchmark}, thus leading to a 7-group comparison.

\subsection{Characterization Methods}

The characteristics of time series, such as variance, seasonality, and autocorrelation, are known to strongly influence the efficacy of change point detection methods~\cite{Wang:09:Characteristics}.
Thus, we begin our investigation by characterizing the performance time series of the three datasets and derive the following characteristics:

\begin{itemize}
    
\item The \textbf{coefficient of variance} (CV) assesses time series stability~\cite{Eismann:22:Serverless}. It gauges performance measurement noise and variance caused by system changes, workload, and non-deterministic factors in tests~\cite{Costa:19:JMH}. CV is calculated as $CV = \frac{\sigma}{\mu} \times 100\%$, where $\sigma$ is the standard deviation and $\mu$ is the mean of the time series.

\item \textbf{Skewness} is a measure of lack of symmetry in the distribution of the time series values. A normally distributed time series has skewness zero. A time series is left skewed if it exhibits a longer tail on the right side of the distribution, and right skewed in the opposite case. Skewness is calculated as $S = \frac{1}{n\sigma^3}\sum_{t=1}^{n}(x_t - \mu)^3$.

\item \textbf{Kurtosis} measures how flat or peaked the data distribution is compared to a normal distribution. To facilitate the interpretation of results, we compute the excess kurtosis when characterizing time series, which normalizes the kurtosis of a normal distribution to zero. The excess kurtosis ($K$) of a time series $X={x_1, x_2, ..., x_n}$ 
is given by $K = \frac{1}{n\sigma^4}\sum_{t=1}{n}(x_t - \mu)^4 - 3$.

\item The \textbf{trend} of a time series helps us characterize its overall trajectory. To compute the trend, we normalize the time series using min-max normalization, and then we compute a third-degree penalized spline with ten knots to extract the average slope~\cite{Eilers1996}.

\item \textbf{Stationarity} measures how time series properties change over time. To test for stationarity, we resort to the Augmented Dickey-Fuller (ADF) test~\cite{Dickey1979}, as used in previous studies~\cite{Wang:09:Characteristics}. ADF tests whether a time series is stationary by checking for a unit root, a stochastic trend in the data that causes it to drift without returning to a constant mean. 
If a unit root is present, the series is non-stationary.

\item \textbf{Autocorrelation} measures the degree of similarity between the time series and a lagged version of itself. We use the Ljung-Box Q-Statistic autocorrelation method to test for the correlation of residuals~\cite{Ljung1978}, as used in previous works~\cite{Wang:09:Characteristics}.

\end{itemize}

To test for statistical differences across seven distributions of time series characteristics, we use the non-parametric Mann-Whitney U test~\cite{wilcoxon}, as we can not assume data normality across time series characteristic distributions.
We report statistical differences assuming a 5\% significance level (p-value < $0.05$).
Furthermore, to quantify the practical impact of the differences, we use the Cliffs-Delta ($d$) effect size~\cite{cliff1993dominance} and resort to the customary interpretation thresholds~\cite{Costa:19:JMH}: 
\textbf{Negligible} (\( |d| < 0.147 \)), \textbf{Small} (\( 0.147 \leq |d| < 0.33 \)), \textbf{Medium} (\( 0.33 \leq |d| < 0.474 \)), and \textbf{Large} (\( |d| \geq 0.474 \)).

\begin{figure*}
    \begin{subfigure}{0.45\textwidth}
        \includegraphics[width=\textwidth]{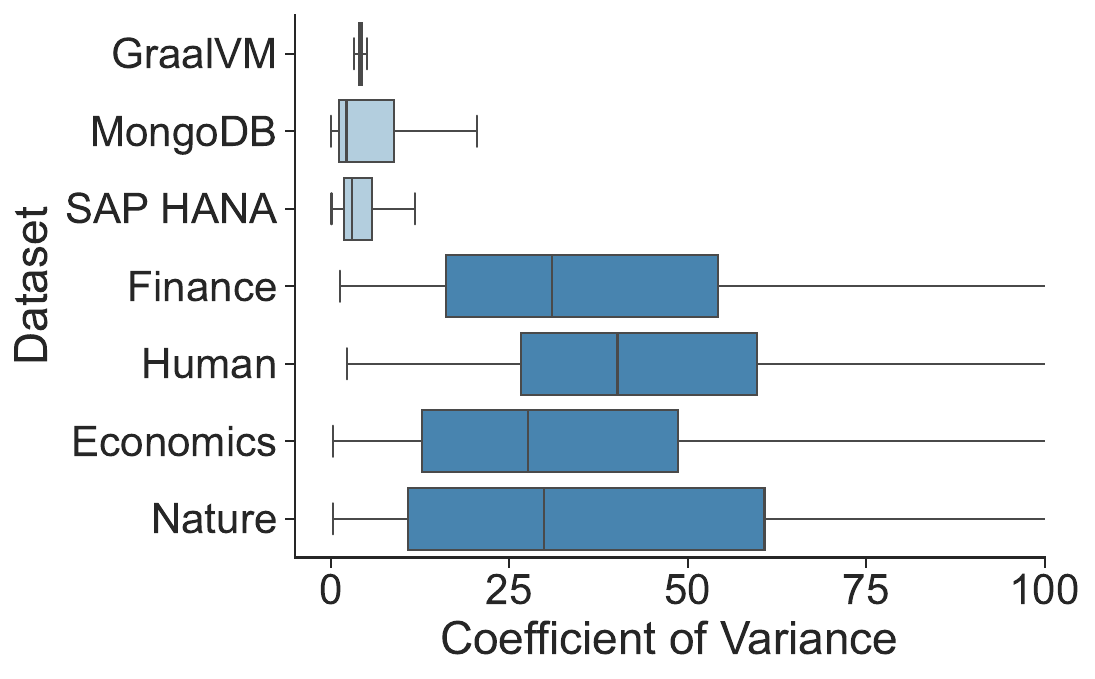}
        \caption{Coefficient of variance.}
        \label{fig:coeffvar_domains}
    \end{subfigure}
    \hfill
    \begin{subfigure}{0.45\textwidth}
        \includegraphics[width=\textwidth]{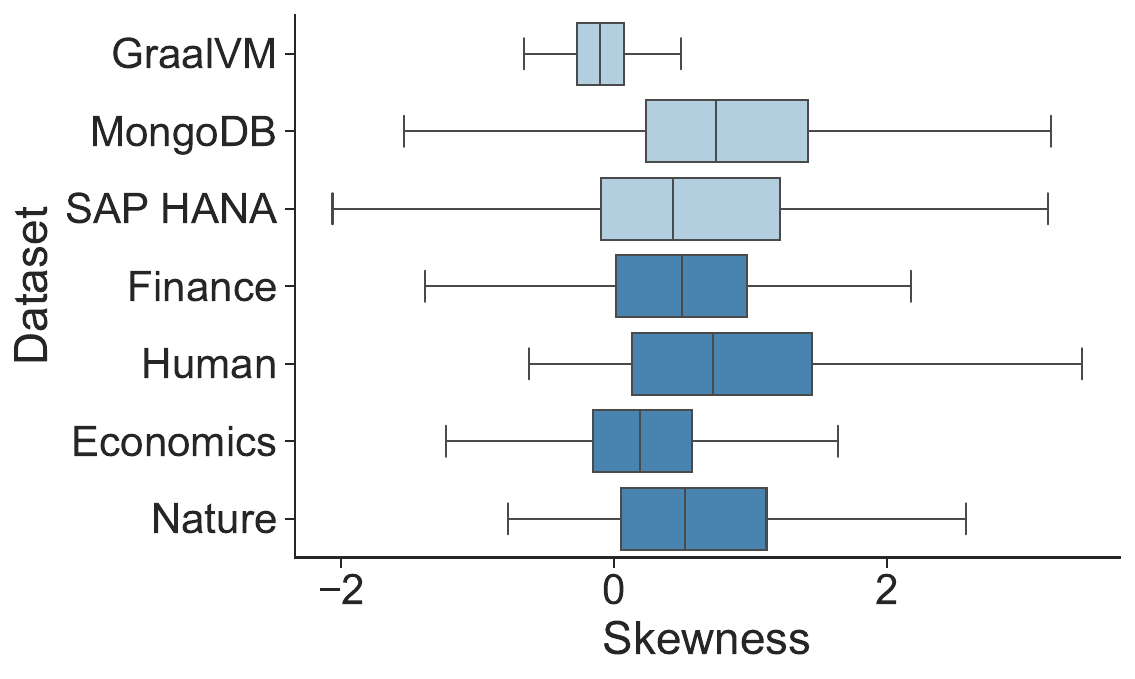}
        \caption{Skewness}
        \label{fig:skewness_domains}
    \end{subfigure}
    \\
    
    \begin{subfigure}{0.45\textwidth}
        \includegraphics[width=\textwidth]{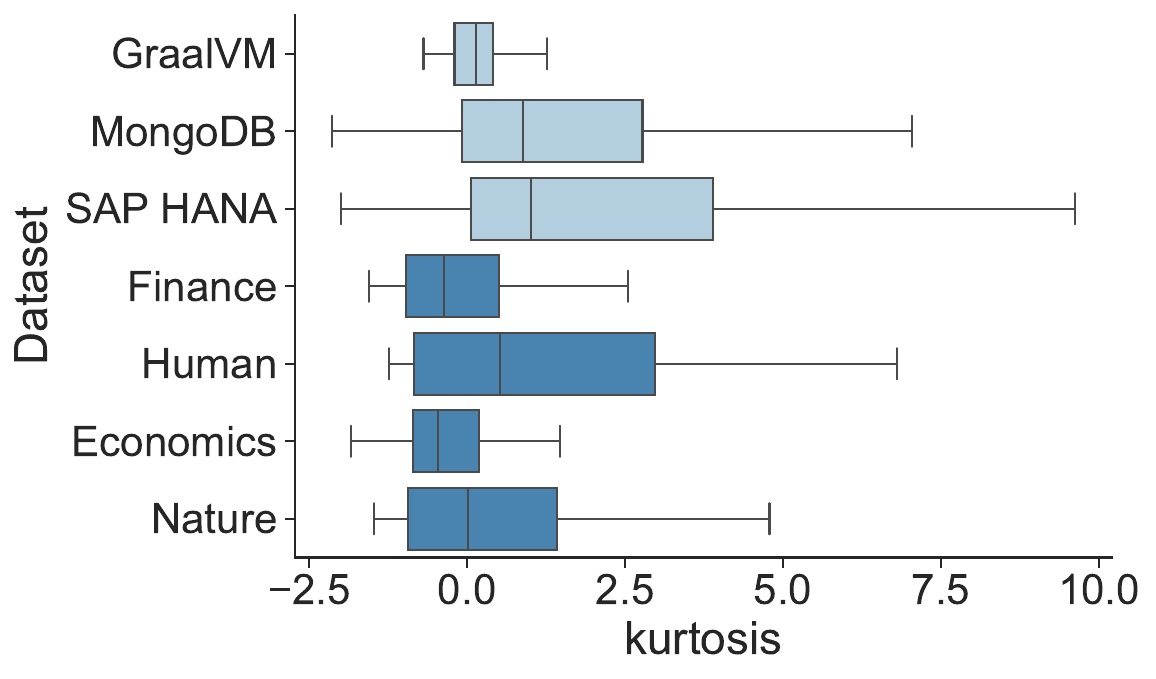}
        \caption{Kurtosis}
        \label{fig:kurtosis_domains}
    \end{subfigure}
    \hfill
    \begin{subfigure}{0.45\textwidth}
        \includegraphics[width=\textwidth]{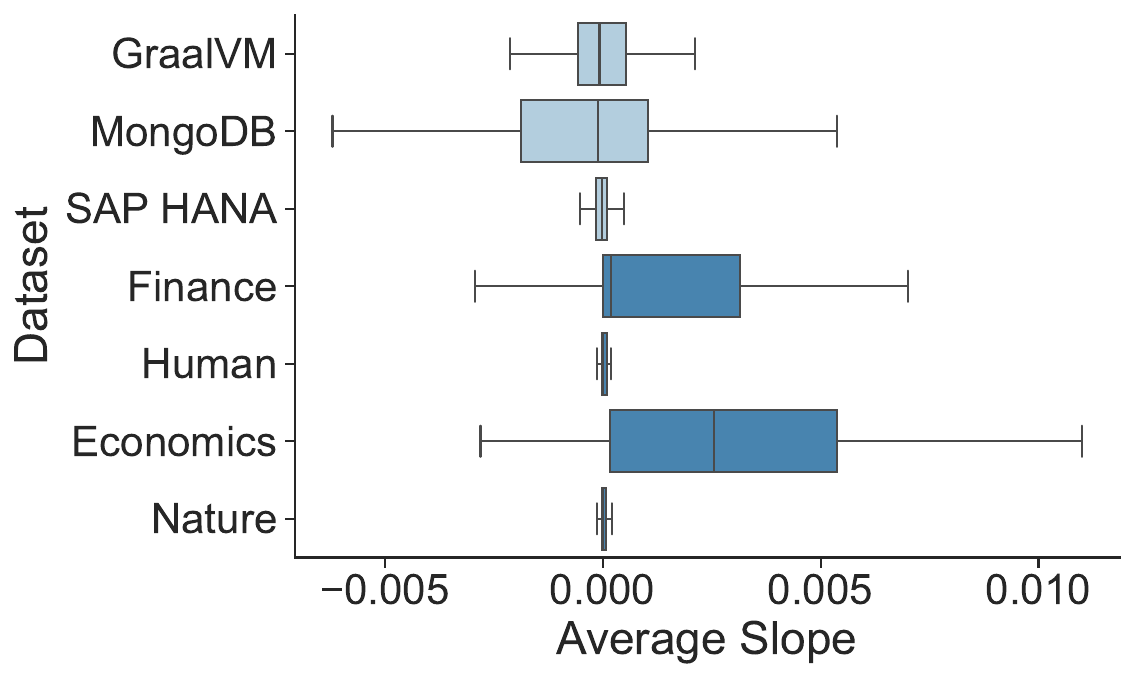}
        \caption{Average Slope (trend)}
        \label{fig:trend_domains}
    \end{subfigure}
    \caption{Comparison of time series characteristics across all datasets (performance and other domains~\cite{Bauer:Libra:benchmark}).}
    \label{fig:comparison_domains}
\end{figure*}

\subsection{RQ1. \rqone}

We present in Figure~\ref{fig:comparison_domains} the comparison of the time series characteristics across our three performance measurement datasets and the four domains from the Libra benchmark~\cite{Bauer:Libra:benchmark}.

\textbf{Finding 1: Performance measurements tend to have less variance and less accentuated trends than the series from other domains, but include more often extreme values (high Kurtosis).} 
When analyzing CoV (Figure~\ref{fig:coeffvar_domains}), we notice that, perhaps counter-intuitively, software performance measurements exhibit a significantly lower coefficient of variance than all the other domains' time series. 
This shows that software performance measurement noise is comparably lower than that of time series from all other domains (large effect size for all comparisons). 
Time series from all other domains tend to exhibit larger variability, with the coefficient of variance commonly reaching 50\% of the mean. 

Regarding skewness (Figure~\ref{fig:skewness_domains}), we found that \graal time series has the least skewed distributions of all the datasets, differing from all other datasets with medium and large effect sizes.
Performance measurements from \sap and \mongo, however, share a similar skewness distribution with time series from Nature, Finance, and Human, with only negligible statistical differences in all pair-wise comparisons (Cliff's delta $|d| < 0.147$). 
Performance measurements are expected to exhibit asymmetrically distributed values (specifically, right-skewness). 
This occurs because system performance has a functional lower bound (minimum execution time) but is subject to various intermittent overheads (e.g., resource contention, context switching, or background processes) that increase latency, creating a long tail of higher values.

When it comes to the presence of outliers (Kurtosis) (Figure~\ref{fig:kurtosis_domains}), we found that \sap and \mongo datasets have a higher presence of outliers compared to all other datasets. 
\graal series, on the other hand, exhibits a tighter distribution with similar kurtosis as shown in Nature and Human datasets.
This is likely due to each point in the time series being derived from an average of many measurements, which tends to reduce its variance. 
Software performance measurements are notorious for producing extreme values~\cite{Costa:19:JMH}, which can pose a challenge for CPD methods.

\begin{figure}
    \begin{subfigure}{0.49\textwidth}
        \includegraphics[width=\textwidth]{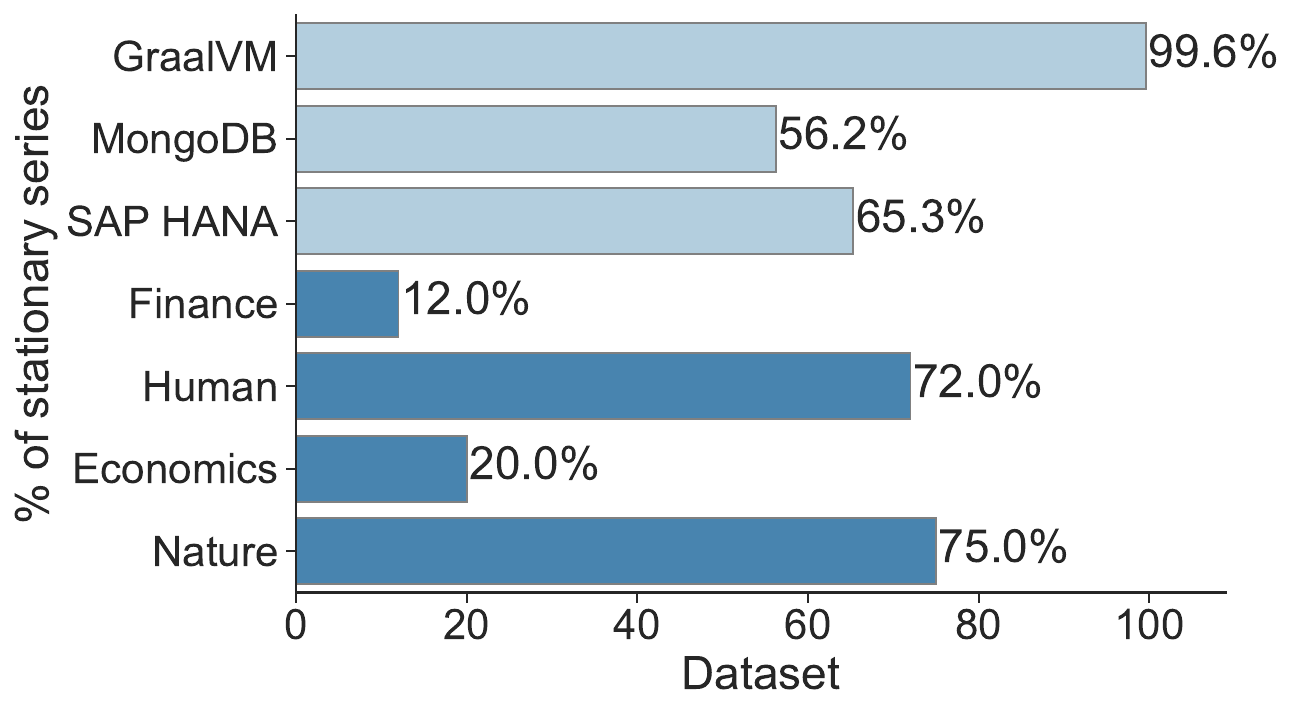}
        \caption{Stationarity}
        \label{fig:stationarity_domains}
    \end{subfigure}
    \hfill
    \begin{subfigure}{0.49\textwidth}
        \includegraphics[width=\textwidth]{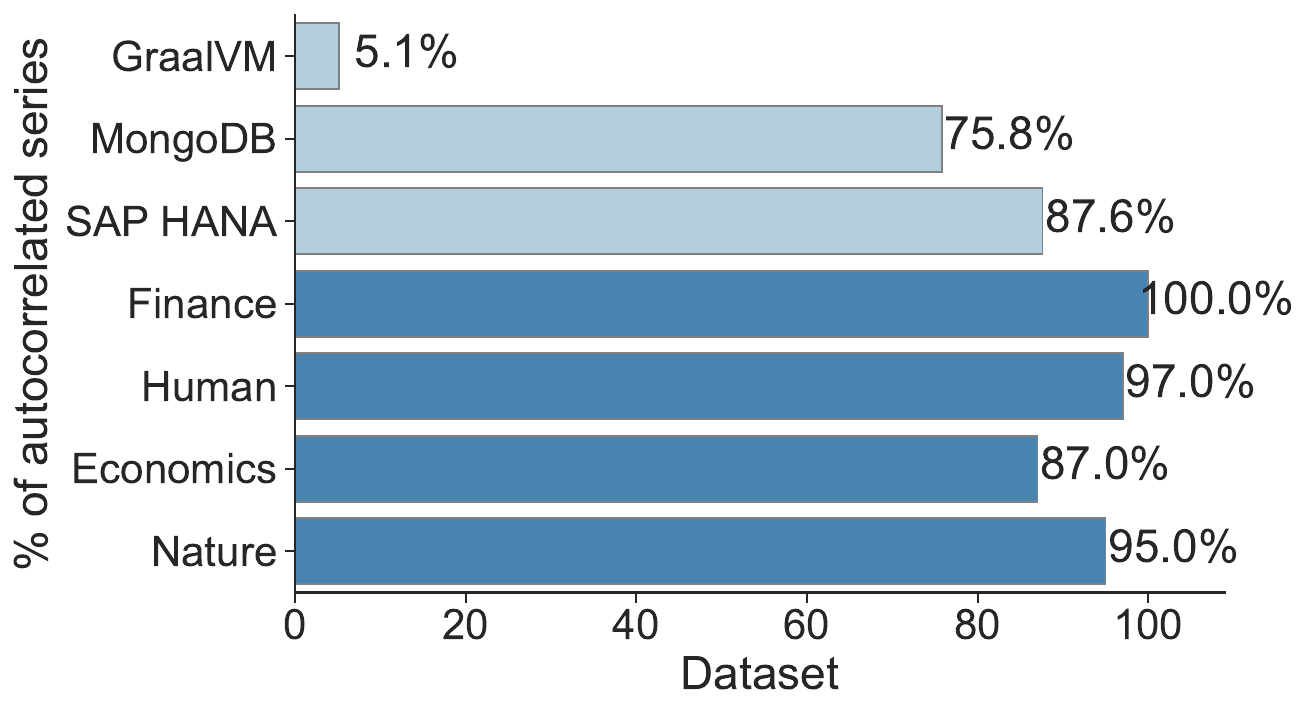}
        \caption{Autocorrelation}
        \label{fig:autocorrelation_domains}
    \end{subfigure}
    \caption{Percentage of stationary and auto-correlated time series across datasets.}
    \label{fig:comparison_stationarity_autocorrelation}
\end{figure}

\textbf{Finding 2: Performance measurements tend to be more stationary but more complex to predict (lower autocorrelation) than series from other domains.}
Figure~\ref{fig:comparison_stationarity_autocorrelation} shows the percentage of stationary and autocorrelated time series in each dataset. 
The differences between the characteristics of performance measurements from \graal against \mongo and \sap are clear. 
\graal is composed primarily of stationary series that are not autocorrelated, indicating performance measurements that are harder to predict (lack of autocorrelation) but that exhibit uniform characteristics (e.g., standard deviation) throughout the measurements. 
\mongo and \sap measurements, on the other hand, have a sizeable share of non-stationary time series, but a higher share of autocorrelation. 
Compared to time series from other domains, all three performance time series exhibit lower autocorrelation, indicating that performance measurements are inherently more stochastic and less reliant on their immediate past values.
This suggests that traditional forecasting models (which depend on strong temporal dependencies) may struggle with software performance data, as the "noise" or high-frequency fluctuations in execution environments obscure the long-term patterns typically found in datasets like finance or nature-related

\section{Annotating suspected Performance Changes}
\label{sec:labeling-approach}

A major challenge for our study is the lack of labelled data to evaluate the accuracy of CPD methods. 
To be effective, a CPD method should distinguish between performance regressions and inherent performance measurement noise. 
In practice, whenever a CPD method raises an alert, a team of performance experts analyze the new measurements and uses their domain' and system's expertise to identify if the alert is worth investigating, which is a process commonly referred to as ``build baroning''~\cite{Daly:21:VirtuousCycle,Daly:20:ChangePointDetection}. %

Given that we do not have access to real system alerts for the datasets, we opt to conduct manual annotation of \textit{potential performance changes}.
We define \textit{potential performance changes} as measurements that, according to a team of performance research experts, should be further investigated by practitioners.  

\subsection{Annotation Tool}\label{sec:annotation-tool}

To perform the annotation of potential performance changes, we use a web-based tool called \textit{AnnotateChange}~\footnote{https://github.com/alan-turing-institute/annotatechange}, developed by Van den Burg \etal as part of their study on change point detection~\cite{Gerrit:2020:ChangePointDetection}.
Upon logging into the tool for the first time, participants must complete a tutorial before they can annotate the time series.
The tutorial consists of several steps in which the user is asked to correctly select change points in different situations, such as when a change in the mean or variance occurs.
Once the users successfully complete the tutorial, the system will select a first time series for them to annotate.
After each annotation task, the next time series to annotate is selected randomly among the ones that are available in the database and have not yet met the required number of annotators.

\begin{figure}
\begin{center}
    \fbox{\includegraphics[width=0.70\linewidth]{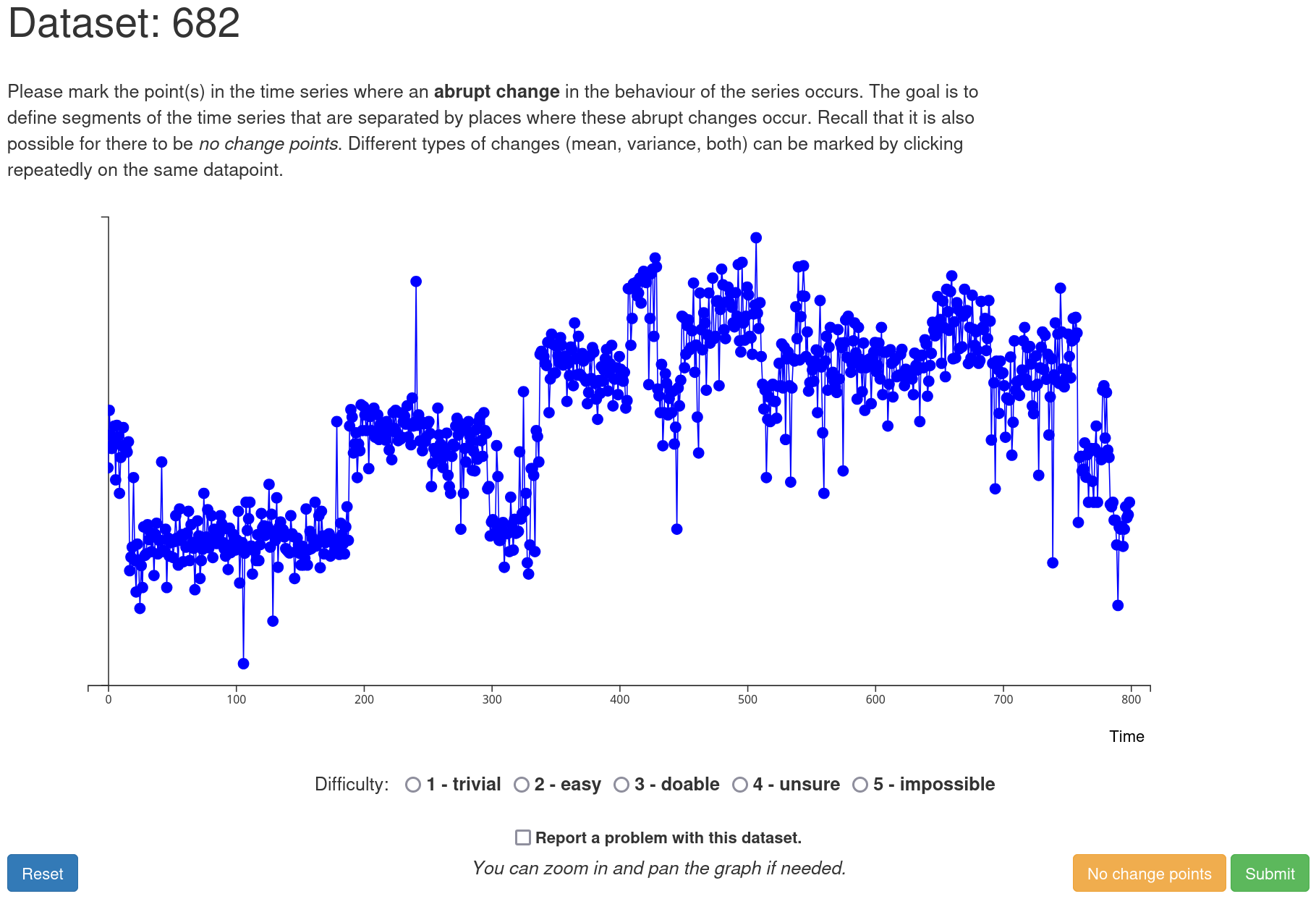}}
\end{center}
\caption{Screenshot of the annotation page of the \emph{AnnotateChange} tool.}
\label{fig:annotation_tool}
\end{figure}

When annotating the performance time series, the participant is presented with a scatter plot of the time series, where data points are connected by straight lines, as shown in \figref{fig:annotation_tool}.
The x-axis is labeled with the consecutive position of the data point in the time series, while the y-axis is not labeled in any way.
The main task of the user is to select the data points representing change points, if any, by clicking on them.
On top of the features that were already provided by the tool, we decided to make some changes and add new features for the purposes of this study:
\begin{itemize}
    \item \textbf{Perceived difficulty:} We included a radio button to let the users grade the difficulty when annotating performance changes into one of five difficulty levels: trivial, easy, doable, unsure, and impossible.
    \item \textbf{Time recorder:} We included a timer to record the time to complete a task when the participant is on the annotation page.
    \item \textbf{Annotation flexibility:} We let the user go back to previously annotated time series to review them and possibly change the annotations and the difficulty preference.
    \item \textbf{Tailored tutorial:} We tailored the tutorial performance time series and introduced the users to different types of annotations. In the tutorial, the users are introduced to the concept of time series change points and have to recognize multiple changes in the mean and variance metrics. 
    Finally, users have to show they can distinguish outliers from change points.
\end{itemize}

\subsection{Pilot Study}

To calibrate the annotation tool and receive preliminary feedback, we first conducted a pilot experiment.  
We manually selected a diverse set of ten performance time series for each of the three datasets. 
After the pilot, we collected feedback from the annotators that helped us in gaining insights about the process and possibly adjusting it.
This also led to further modify the annotation tool accordingly.

The annotators who participated in the pilot study reported 493 change points distributed over 30 time series.
We noticed that, on average, an annotator spent approximately one minute for each time series.
Most users reported that the difficulty setting was not visible enough in the user interface of the annotation page, and this made them forget about selecting a difficulty level in some cases.
Hence, we modified the user interface accordingly, and we prevented the tool to proceeding to the next time series when the difficulty level is not set.
In the new version of the tool, for each time series, we collect the annotated data points, the difficulty preference, and the time spent annotating.
The newly developed features are publicly available\footnote{\url{https://github.com/mtucci/AnnotateChange}}.

\subsection{Sampling the Performance Time Series}
Our dataset contains more than 20 thousand performance time series. 
Given that analyzing and annotating real performance time series is a time-consuming task, we sample our datasets to include a target of 500 time series per dataset, 1500 in total. 
We performed a stratified sampling by accounting for the number of potential change points yielded by the CPD methods~\cite{baltes2022sampling}.

\begin{itemize}
    \item We run all the CPD methods and compute the number of unique change points yielded by the set of methods.
    \item We then randomly select 100 performance time series from each 20th percentile of the number of unique change point distributions. 
\end{itemize}

This method ensures proportional group representation, where the group of stable time series ($<$ 20th percentile) will have equal representation in the final sample as groups with a higher number of potential performance changes.

\subsection{Annotating Performance Changes}

Our goal with the annotation process is to establish a high-quality dataset of potential performance changes built on real performance measurements. 
To achieve this, we first define the workload expected from participants and recruit participants using a convenience sample methodology~\cite{baltes2022sampling}.

\textbf{Participant Selection and Workload.} 
We opt to include only participants, hereby named as annotators, with experience in performance engineering in industry or research. 
In total, we selected 14 annotators: 10 are authors of this paper, and 4 were external annotators recruited for this study. The demographics and experience of the annotators are presented in Table~\ref{tab:demographics}. %

\textbf{Annotation Workload.}
We tasked each annotator to label 720 performance time series, estimating six to seven hours of work. 
The annotation tool was configured to allow a maximum of five annotators per time series, but to prioritize distributing partially labelled time series over non-annotated ones.
We gave participants four weeks to complete the annotation process and compile the results after this period.

\begin{table}[]
    \centering
    \caption{Demographics of the 14 annotators.}
    \label{tab:demographics}
    \begin{tabular}{l l r}
        \toprule
          & \textbf{Category} & \textbf{\#} \\
         \midrule
         
        \multirow{3}{*}{\textbf{Profession}} 
        & Performance Engineer & 4 \\    
        & Assistant/Associate/Full Professor & 7 \\
        & Research Associate & 3 \\    

        \midrule
        \multirow{4}{*}{\textbf{Experience}} 
         & 20+ years & 3 \\
         & 10-20 years & 2 \\
        & 5-10 years & 5 \\
        & 0-5 years & 4 \\

         \bottomrule
    \end{tabular}

\end{table}

\subsection{Evaluating Annotation agreement}
\label{sec:evaluation-agreement}

After annotating the performance change points in our sampled set of 1500 performance time series, we now move towards evaluating the agreement between the annotators.
High levels of agreement between annotators may indicate a level of robustness in the reported change points. 
On the other hand, a significant divergence in the change points identified by annotators implies a lack of consensus, suggesting that the results could be unstable and prone to subjectivity.

\noindent
\textbf{Filtering series with fewer than five annotators.} 
To keep our analysis consistent, we first exclude time series annotated by fewer than five annotators, thus excluding 343 series that did not reach the five annotators. 
We are left with 1154 time series fully annotated by five annotators, 380 from \graal, 383 from \mongo, and 391 from \sap. 
Our final dataset comprises 10028 change point annotations identified by 15 distinct annotators.

\noindent
\textbf{Agreement metrics.}
To quantitatively assess the agreement of the annotators, we evaluate each individual annotator against the four others on every time series, using the $F_1$-score (equation (\ref{eq:f1})).
We refer to this metric as one-vs-rest (OvR) $F_1$-score.
A OvR $F_1$-score close to 1 indicates high agreement, because the annotators identified similar sets of change points on a given time series, while a value close to 0 implies substantial disagreement among annotators.

\noindent
\textbf{Perceived Difficulty.}
We also assess the perceived difficulty of the annotators during the detection of performance changes, and explore how this perception correlates with their level of agreement (\ie OvR $F_1$-score). 
Lastly, we study the duration of each labeling task to gain insight into the time required for manually identifying performance changes.

\subsection{RQ2: \rqtwo}

\begin{table}[]
    \centering
    \caption{Statistics on the number of \potentialchangepoints found during our annotation experiment.
    }
    \label{tab:annotation_stats}
    \begin{tabular}{lrr}
        \toprule
         Dataset & \# Series & Avg \# of \potentialchangepoints\\
         \midrule   
         \mongo & 383 & 2.5  \\
         \graal & 380 & 1.1 \\
         \sap & 391 & 3.2 \\
         \bottomrule
    \end{tabular}

\end{table}

Table \ref{tab:annotation_stats} presents statistics on the suspected change points identified by annotators for each time series across the three datasets. On average, individual annotators identified 2.5, 1.1, and 3.2 change points per time series in the \mongo, \graal, and \sap datasets, respectively.

\begin{figure}
    \centering
    \includegraphics[width=\linewidth]{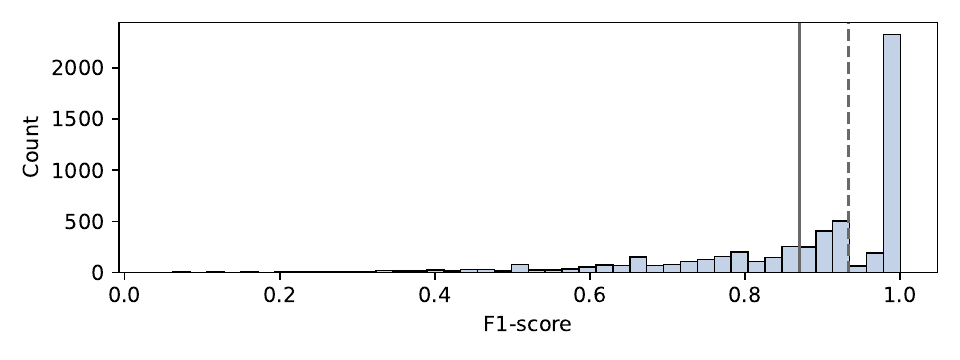}
    \caption{The histogram depicts the distribution of the OvR $F_1$-score metric for every annotator across all performance time series. The solid vertical line represents the mean, while the dashed line denotes the median.}
    \label{fig:ovr}
\end{figure}

\textbf{Finding 1: Annotators highly agree on \potentialchangepoints, with most performance series (74\%) achieving an OvR $F_1$-score greater than 0.8.}
Figure~\ref{fig:ovr} shows the distribution of the OvR $F_1$-score metric across all annotation tasks. The distribution highlights a high level of agreement among annotators. As illustrated in \figref{fig:ovr}, annotators achieve a mean OvR $F_1$-score of 0.87 and a median of 0.93 (IQR 0.8-1.0). Notably, about 40\% of the annotation tasks result in a perfect OvR $F_1$-score of 1.0.
These results indicate a convergence in the change points labeled by different annotators, suggesting a substantial agreement among them.

\begin{figure}
    \includegraphics[width=.98\linewidth]{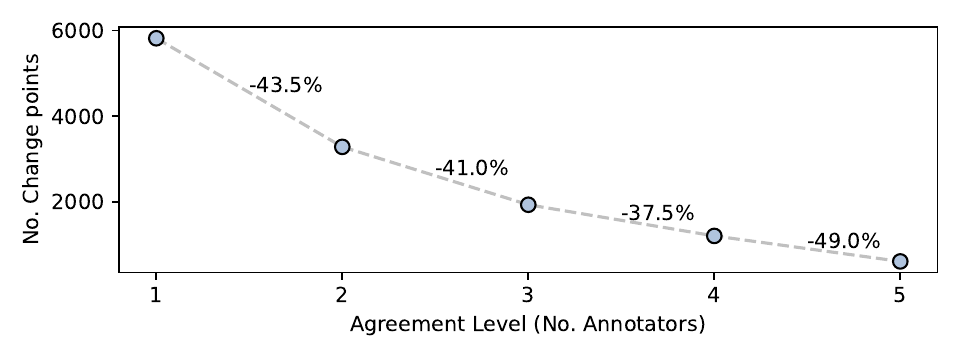}
\caption{Number of detected change points under different agreement levels. Each circle represents the number of change points identified by at least $n$ annotators, with $1\leq n \leq 5$.} 
\label{fig:agreement-sensitivity}
\end{figure}

\textbf{Finding 2: Despite exhibiting high agreement, identifying potential performance changes may remain susceptible to subjective interpretation.}
\figref{fig:agreement-sensitivity} shows the number of unique change points\footnote{Unique change points are assessed with a margin of error \( M=5 \). 
For instance, if two different change points are identified by two distinct annotators and are within 5 subsequent data points within the time series, they are counted as one.} that are identified by at least \( n \) annotators, where \( n \) ranges from 1 to 5. 
As illustrated by the figure, the number of change points considerably decreases when multiple annotators are considered.
For instance, the left-most circle denotes a total of 5,815 unique change points labeled by at least one annotator, while the subsequent circle indicates 3,286 unique change points identified by at least two annotators, highlighting a significant decrease of -43.5\%.
This downward trend continues when considering the change points identified by at least three, four, and five annotators, resulting in 1,938 (-41.0\%), 1,212 (-37.5\%), and 618 (-49\%) unique change points, respectively.
These results indicate subjectivity in performance change detection, suggesting that some annotators consider certain change points as notable performance shifts, while others do not.

These observations seem to contrast with our earlier findings, which underscored a broad agreement among annotators; however, there is a subtle aspect to be considered here. 
The notion of agreement, along with the associated OvR metrics utilized in this study, is based on the change points identified by multiple annotators, thereby delivering an ``averaged'' assessment that mitigates the threat of subjective evaluations.
In that, it becomes apparent that relying solely on one annotator might not always yield reliable results.
On the other hand, our results also indicate that the identification of change points tends to converge when averaged over the viewpoints of multiple annotators, as confirmed by the high OvR scores achieved by the annotators.

\begin{figure}
    \includegraphics[width=.98\linewidth]{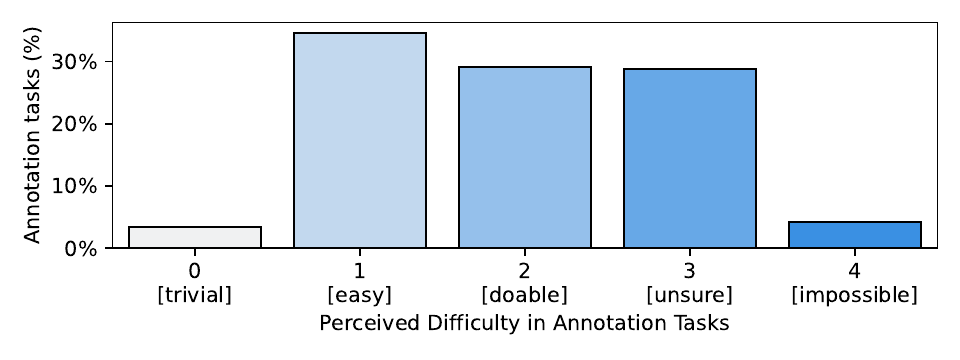}
\caption{Perceived difficulty in annotation tasks. The bar chart report the distribution of the perceived difficulty in each annotation task. Difficulty is assessed using a 5-point Likert scale, where 1 indicates low difficulty and 5 denotes high difficulty.}
\label{fig:difficulty}
\end{figure}

\textbf{Finding 3: Annotating \suspiciouschangepoints was doable for 67\% of the series, but in 5\% of the time series annotators found it impossible to distinguish noise from changes}.  
\figref{fig:difficulty} shows the percentages of perceived difficulty in annotation tasks.
As illustrated in the figure, only 3.5\% of the tasks have been considered as \emph{trivial} by the annotators. 34.5\% have been perceived as \emph{easy} and 29\% as \emph{doable}.
Interestingly, during their annotation tasks a notable 28.7\% of annotators were \emph{unsure} about the performance shifts observed in the time series, while only 4.2\% of the tasks were deemed as \emph{impossible}. 
The results suggest that identifying performance shifts in real-world metrics is often perceived as feasible: specifically, about 67\% of tasks were labeled as either \emph{trivial}, or \emph{easy} or \emph{doable}. Nonetheless, a notable 33\% of the tasks were considered as either \emph{unsure} or \emph{impossible}, highlighting challenges in the analysis of the time series.

\begin{figure}
    \centering
        \includegraphics[width=\linewidth]{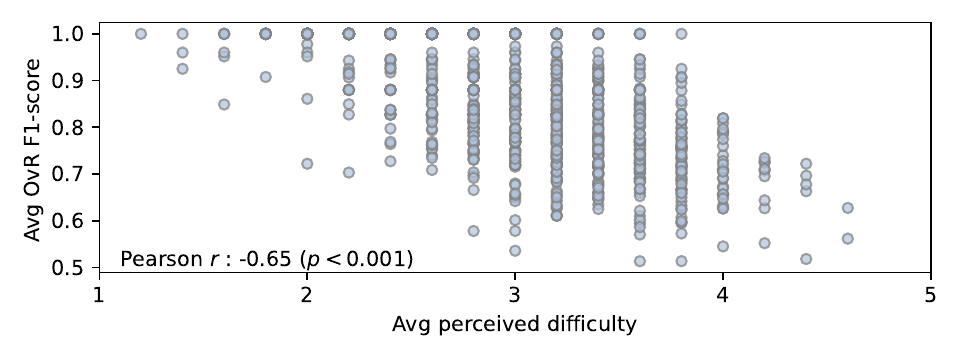}
\caption{Difficulty \emph{vs} OvR $F_1$-score.}
\label{fig:difficulty-vs-f1}
    \label{fig:difficulty-vs-ovr}
\end{figure}

To better understand these factors, we assess the correlation between the agreement (\ie OvR $F_1$-score metric) and the difficulty reported by the annotators.
\figref{fig:difficulty-vs-ovr} shows the relationship between these two factors, where the y-axis displays the average OvR $F_1$-score that has been achieved by annotators, and the x-axis represents the average difficulty perceived by the annotators. 
Each circle on the scatter plot refers to a specific time series.
As illustrated in the figure, and as expected, the OvR score appears to be inversely correlated with the difficulty score.
To further investigate this aspect, we calculated the Pearson correlation coefficient ($r$), finding a moderate negative correlation of $r=-0.65$ in the OvR $F_1$-score metric with strong statistical significance (p-value $<$ 0.001).
These findings suggest a likely higher disagreement in time series perceived as more difficult to analyze.
A possible explanation for this behavior is that as the analysis of time series becomes more difficult, the identification of performance changes becomes more prone to subjective interpretation and, therefore, to higher disagreement among annotators.
This finding suggests that using multiple annotators is especially important when dealing with challenging time series, as the variability in change point identification may substantially increase in such cases.

\begin{figure}
    \includegraphics[width=.98\linewidth]{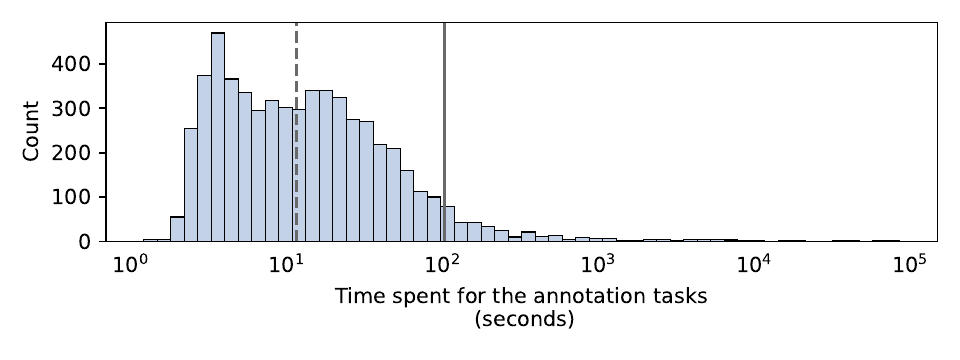}
\caption{Distribution of the time spent for each annotation task in seconds. The solid vertical line represents the mean, while the dashed line denotes the median.}
\label{fig:time}
\end{figure}

\begin{figure}
    \includegraphics[width=.98\linewidth]{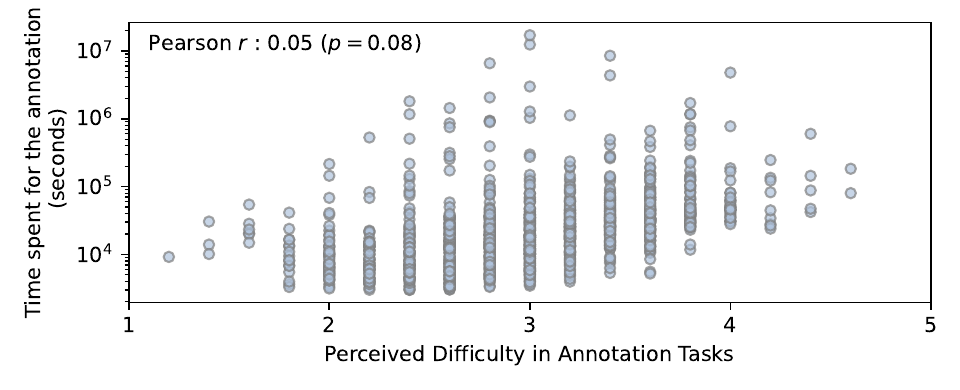}
\caption{Difficulty \emph{vs.} Annotation time}
\label{fig:difficulty-time}
\end{figure}

Our study further reports on the duration of the performance changes identification process.
\figref{fig:time} depicts the distribution of the time required for each annotation task. %
As illustrated in the figure, the annotation took an average duration of 101 seconds and a median of 11 seconds. 
While these durations might seem negligible at first glance, they become significant when dealing with complex performance assurance processes that involve a large number of performance tests (see, for example, \cite{Daly:21:VirtuousCycle, Traini2022}).
For a practical example, let us consider MongoDB's performance testing suite, which involves thousands of tests \cite{Daly:21:VirtuousCycle}. At the mean time of 101 seconds per test, annotating the entire suite would require more than three days of work.
Moreover, the necessity of involving multiple annotators for gathering accurate results (especially when performance changes are challenging to pinpoint) further amplifies this time commitment.
Figure \ref{fig:difficulty-time} also shows a slight relationship between the difficulty of the time series and the correlation time, although the correlation is not statistically significant (Pearson $r$ = 0.05, $p$ = 0.08).

Overall, such findings underscore the potential benefits and need for transitioning to automated performance change detection methods. %

\section{Evaluating Change Detection Methods}
\label{sec:change-detection-methods}

In this section, we evaluate a large selection of major CPD algorithms from the study of van den Burg and Williams~\cite{Gerrit:2020:ChangePointDetection}.

\begin{table*}
\centering
\caption{The selected change point detection algorithms.}
\label{tab:algo}
    \resizebox{\textwidth}{!} {
\begin{tabular}{l l l r}
\toprule
Name & Method and Refs & Parameters & \# Combinations \\
\midrule
AMOC & At Most One Change~\cite{Hinkley1970} & function, Q, penalty, statistic & 126  \\

BINSEG & Binary Segmentation~\cite{Scott1974} & function, Q, penalty, statistic & 252 \\

BOCPD & Bayesian Online Change Point Detection~\cite{Adams2007} & prior\_k, prior\_b, prior\_a, intensity &  81 \\

CPNP & Nonparametric Change Point Detection~\cite{Haynes2017} & penalty, quantiles & 28 \\

ECP & Energy Change Point~\cite{Matteson2014} & alpha, algorithm, siglvl, minsize & 24 \\

KCPA & Kernel Change-Point Analysis~\cite{Harchaoui2008}  & maxcp, cost & 14 \\

MONGO & CPD Method by MongoDB performance team~\cite{Daly:20:ChangePointDetection} & pvalue, permutations & 12 \\

PELT & Pruned Exact Linear Time~\cite{Killick2012} & function, Q, penalty, statistic & 126 \\

PROPHET & Prophet~\cite{Taylor2018} & nmax & 2 \\

RFPOP & Robust Functional Pruning Optimal Partitioning~\cite{Fearnhead2019} & loss & 4 \\

SEGNEIGH & Segment Neighborhoods~\cite{Auger1989} & function, Q, penalty, statistic & 252 \\

WBS & Wild Binary Segmentation~\cite{Fryzlewicz2014} & Kmax, integrated, penalty & 12 \\  
\bottomrule
\end{tabular}
}
\end{table*}

\subsection{Change Point Detection Methods}

Inspired by the benchmark study presented by \father, we aim to determine if automated change point detection methods can perform at a level reasonably close to human annotators.
Most of these approaches are offline change point detection methods, which assume access to the complete time series and dominate the existing CPD benchmarking literature~\cite{Truong2020}. In contrast, online methods process observations sequentially as new data arrive. Among the evaluated techniques, BOCPD~\cite{Adams2007} represents this class and is included as a reference point for online detection. Even if in the performance domain, online methods provide the advantage of processing observations sequentially and update their estimates as new data points (new software measurements for new versions) arrive, we focus primarily on offline methods to enable a consistent comparison across algorithms on the same complete time series while still providing insight into how an online approach performs in this context.
To this extent, we executed a selection of $12$ methods that are open-source, actively maintained, and distributed as readily available software packages.
Table~\ref{tab:algo} lists the methods included in the study.
Each method included a series of parameters that can be tuned, and their combinations vary from 2 to 252. 

\rev{R2C4}{
In addition to the methods evaluated in \father, we also included the \texttt{MONGO} method, which is an adaptation of the E-Divisive Means approach adopted by MongoDB for performance regression detection~\cite{Daly:20:ChangePointDetection}.
The method was selected by Daly et al. because it is non-parametric, supports the detection of multiple change points, and does not require training data. We use the Python implementation released by MongoDB, with \texttt{pvalue} and \texttt{permutations} as tunable parameters.
}

We report the effectiveness of the methods under two settings: 

\begin{itemize}
    \item \textbf{Default}, where default values were used for all the parameters to evaluate how CPD methods perform in an out-of-the-box fashion. 

    \item  \textbf{Oracle}, \rev{R2C2}{where an exhaustive grid search was performed and, for each time series, we selected the parameter configuration yielding the highest $F_1$-score.}
\end{itemize}

We computed the effectiveness of the selected methods using the metrics described in Section~\ref{sub:evaluation-metrics}.
\rev{R2C2}{
Specifically, in the Oracle setting, we exhaustively evaluated on each time series all the combinations of parameters reported in Table~\ref{tab:algo}.
For each method and time series, we selected the highest $F_1$-score obtained across all parameter configurations. The Oracle $F_1$-score reported for a method is the average of these best per-series scores. Precision and recall were taken only from configurations achieving the best $F_1$-score for that series. When several configurations achieved the same best $F_1$-score, we reported the highest precision and the highest recall among them.
Therefore, the Oracle setting assumes knowledge of the ground truth for each individual time series and represents an upper bound on the performance attainable through hyperparameter tuning rather than a tuning strategy that can be applied in practice.
}

\subsection{Evaluation Metrics}
\label{sub:evaluation-metrics}

The effectiveness of a CPD method can be quantified in terms of precision (the ratio of correctly detected change points over the number of detected change points) and recall (the ratio of correctly detected change points over the number of true change points).
We use the \emph{$F_1$-score}, \ie the harmonic mean of precision ($P$) and recall ($R$), which provides a unique measure to evaluate the quality while keeping into consideration both these aspects.

$F_1$-score quantifies the accuracy of a method by considering CPD as a classification problem, where each data point of the series can be either classified as ``change point'' and ``non-change point''.

As a result of our manual labeling process (described in Section~\ref{sec:labeling-approach}), we obtain (for each time series) multiple annotations provided by different annotators.
In particular, the locations of change points provided by the annotator $k\in \{1,...,K\}$ are denoted by the ordered set $\mathcal{T}_k = \{\tau_1,...,\tau_{n_k}\}$, with $1\leq\tau_i\leq T$ and $T$ denoting the length of the time series.

In order to reduce the amount of mismatches, in CPD, it is common to define a margin of error $M$ around the true change point location \cite{Killick2012,Truong2020}.
Hence, to handle the multiple sets of ground truth annotations, we follow the approach proposed by Martin et al.~\cite{Martin2004}.
Given a set of change point locations $\mathcal{X}$ provided by a detection method and a set of human annotations $\mathcal{T}$, we define the set of true positives $TP(\mathcal{T}, \mathcal{X})$ as the set of $\tau\in\mathcal{T}$ for which $\exists x\in \mathcal{X}$ such that $|\tau-x|\leq M$. This process ensures that only one $x\in\mathcal{T}$ can be used for a single ground-truth change point location $\tau\in\mathcal{T}$.

Based on this definition, precision and recall can be computed as follows:

\begin{equation}
	P = \frac{|TP(\mathcal{T}^*, \mathcal{X})|}{|\mathcal{X}|}	
\end{equation}

\begin{equation}
	R = \frac{1}{K}\sum_{k=1}^{K}\frac{|TP(\mathcal{T}_k, \mathcal{X})|}{|\mathcal{T}_k|}	
\end{equation}
where $\mathcal{T}^*$ denotes the whole set of human annotations $\bigcup_{k}\mathcal{T}_k$.

And $F_1$-score can be computed accordingly, using the usual formula~\cite{VanRijsbergen1979}:

\begin{equation}
	F_1=\frac{2PR}{P+R}
	\label{eq:f1}
\end{equation}

\subsection{RQ3. \rqthree}

\begin{table*}
    \centering
    \caption{Evaluation results of our CPD experiment. We include the average results across datasets (``Overall'' column) and the results per dataset on both the \textit{Default} and \textit{Oracle} experiments. We highlight the methods that achieved the best $F_1$-scores in each dataset.}
    \label{tab:scores_overall_combined}
    
\definecolor{heatw}{RGB}{255, 255, 255} %
\definecolor{heat0}{RGB}{230, 245, 255}  %
\definecolor{heat1}{RGB}{200, 230, 253}  %
\definecolor{heat2}{RGB}{170, 215, 251}  %
\definecolor{heat3}{RGB}{135, 206, 250}  %
\definecolor{heat4}{RGB}{100, 185, 235}  %
\definecolor{heat5}{RGB}{65, 160, 220}   %
\definecolor{heat6}{RGB}{45, 130, 200}   %
\definecolor{heat7}{RGB}{25, 100, 180}   %
\definecolor{heat8}{RGB}{15, 75, 150}    %
\definecolor{heat9}{RGB}{5, 50, 120}     %

\newcommand{\heatmap}[3]{%
  \pgfmathparse{(#3-#1)/(#2-#1)}%
  \ifdim\pgfmathresult pt<0.20pt
    \cellcolor{heatw}#3%
  \else\ifdim\pgfmathresult pt<0.40pt
    \cellcolor{heat0}#3%
  \else\ifdim\pgfmathresult pt<0.60pt
    \cellcolor{heat1}#3%
  \else\ifdim\pgfmathresult pt<0.80pt
    \cellcolor{heat2}#3%
  \else\ifdim\pgfmathresult pt<1.00pt
    \cellcolor{heat3}#3%
   \else
    \cellcolor{heat4}#3%
  \fi\fi\fi\fi\fi
}

\newcommand{\heatmapCPD}[1]{\heatmap{0.9}{1.00}{#1}}

\newcommand{\gold}{\cellcolor{heat4}}
\newcommand{\silver}{\cellcolor{heat2}}
\newcommand{\bronze}{\cellcolor{heat0}}

\begin{adjustbox}{width=\linewidth}
\begin{tabular}{l|rrr|rrr|rrr|rrr|rrr|rrr|rrr|rrr}
    \toprule

    & \multicolumn{6}{c|}{\textbf{Overall}}
    & \multicolumn{6}{c|}{\textbf{\graal}}
    & \multicolumn{6}{c|}{\textbf{\mongo}}
    & \multicolumn{6}{c}{\textbf{\sap}}
    \\

    & \multicolumn{3}{c|}{\textbf{Default}} & \multicolumn{3}{c|}{\textbf{Oracle}} %
    & \multicolumn{3}{c|}{\textbf{Default}} & \multicolumn{3}{c|}{\textbf{Oracle}} %
    & \multicolumn{3}{c|}{\textbf{Default}} & \multicolumn{3}{c|}{\textbf{Oracle}} %
    & \multicolumn{3}{c|}{\textbf{Default}} & \multicolumn{3}{c}{\textbf{Oracle}} %
    \\
    Method 
        & $P$ & $R$ & $F_1$
        & $P$ & $R$ & $F_1$ %
        
        & $P$ & $R$ & $F_1$ 
        & $P$ & $R$ & $F_1$ %
        
        & $P$ & $R$ & $F_1$
        & $P$ & $R$ & $F_1$ %
        
        & $P$ & $R$ & $F_1$
        & $P$ & $R$ & $F_1$ %
    
    \\
    \midrule

AMOC
    & 0.97 & 0.79 & 0.84 & 1.00 & 0.85 & 0.90 %
    & 1.00 & 0.95 & \gold 0.97 & 1.00 & 0.99 & \gold 0.99 %
    & 0.98 & 0.67 & \bronze 0.77 & 1.00 & 0.74 & 0.83 %
    & 0.92 & 0.74 & \bronze 0.79 & 1.00 & 0.77 & 0.85 %
    \\

BINSEG
    & 0.94 & 0.83 & \gold 0.87 & 1.00 & 0.95 & \gold 0.97 %
    & 1.00 & 0.95 & \gold 0.97 & 1.00 & 0.98 & \gold 0.99 %
    & 0.95 & 0.75 & \gold 0.82 & 0.99 & 0.93 & \silver 0.95 %
    & 0.89 & 0.78 & \gold 0.81 & 0.99 & 0.90 & \gold 0.94 %
    \\

BOCPD
    & 0.80 & 0.91 & 0.82 & 0.99 & 0.95 & \gold 0.97 %
    & 0.99 & 0.96 & \gold 0.97 & 1.00 & 0.96 & \silver 0.98 %
    & 0.79 & 0.87 & \silver 0.80 & 0.99 & 0.94 & \gold 0.96 %
    & 0.62 & 0.91 & 0.69 & 0.97 & 0.92 & \silver 0.93 %
    \\

CPNP
    & 0.69 & 0.92 & 0.75 & 0.75 & 0.95 & 0.81 %
    & 0.77 & 0.97 & 0.82 & 0.79 & 0.97 & 0.84 %
    & 0.71 & 0.90 & 0.76 & 0.79 & 0.95 & 0.84 %
    & 0.58 & 0.91 & 0.67 & 0.64 & 0.93 & 0.72 %
    \\

ECP
    & 0.84 & 0.85 & 0.81 & 0.96 & 0.94 & \bronze 0.94 %
    & 0.99 & 0.95 & \gold 0.97 & 1.00 & 0.96 & \silver 0.98 %
    & 0.81 & 0.73 & 0.73 & 0.97 & 0.91 & 0.93 %
    & 0.71 & 0.86 & 0.74 & 0.89 & 0.93 & 0.89 %
    \\

KCPA
    & 0.03 & 1.00 & 0.05 & 0.86 & 0.90 & 0.86 %
    & 0.01 & 1.00 & 0.03 & 0.99 & 0.96 & \bronze 0.97 %
    & 0.06 & 1.00 & 0.10 & 0.81 & 0.89 & 0.83 %
    & 0.01 & 1.00 & 0.03 & 0.76 & 0.86 & 0.77 %
    \\

MONGO
    & 0.88 & 0.88 & \silver 0.86 & 0.93 & 0.92 & 0.91 %
    & 0.99 & 0.95 & \gold 0.97 & 0.99 & 0.96 & \bronze 0.97 %
    & 0.86 & 0.83 & \gold 0.82 & 0.92 & 0.87 & 0.88 %
    & 0.79 & 0.85 & \silver 0.80 & 0.84 & 0.90 & 0.85 %
    \\

PELT
    & 0.89 & 0.85 & \bronze 0.84 & 0.97 & 0.95 & \silver 0.95 %
    & 1.00 & 0.95 & \gold 0.97 & 1.00 & 0.97 & \gold 0.99 %
    & 0.92 & 0.78 & \gold 0.82 & 0.96 & 0.93 & \bronze 0.94 %
    & 0.77 & 0.82 & 0.74 & 0.87 & 0.92 & 0.86 %
    \\

PROPHET
    & 0.77 & 0.71 & 0.69 & 0.91 & 0.74 & 0.79 %
    & 0.98 & 0.95 & \silver 0.96 & 1.00 & 0.95 & \bronze 0.97 %
    & 0.72 & 0.54 & 0.56 & 0.87 & 0.55 & 0.63 %
    & 0.63 & 0.65 & 0.57 & 0.85 & 0.68 & 0.72 %
    \\

RFPOP
    & 0.73 & 0.94 & 0.78 & 0.83 & 0.96 & 0.86 %
    & 0.96 & 0.96 & \bronze 0.95 & 0.98 & 0.97 & \bronze 0.97 %
    & 0.60 & 0.94 & 0.68 & 0.67 & 0.96 & 0.75 %
    & 0.64 & 0.93 & 0.72 & 0.71 & 0.94 & 0.77 %
    \\

SEGNEIGH
    & 0.49 & 0.89 & 0.57 & 0.99 & 0.95 & \gold 0.97 %
    & 0.27 & 0.98 & 0.41 & 1.00 & 0.97 & \gold 0.99 %
    & 0.66 & 0.87 & 0.71 & 0.98 & 0.92 & \bronze 0.94 %
    & 0.55 & 0.84 & 0.59 & 0.99 & 0.89 & \bronze 0.92 %
    \\

WBS
    & 0.62 & 0.95 & 0.68 & 0.80 & 0.94 & 0.83 %
    & 0.96 & 0.96 & \bronze 0.95 & 0.99 & 0.96 & \bronze 0.97 %
    & 0.39 & 0.95 & 0.48 & 0.55 & 0.94 & 0.62 %
    & 0.53 & 0.93 & 0.62 & 0.71 & 0.92 & 0.77 %
    \\

    \midrule

\textbf{Average} 
    & 0.72 & 0.88 & 0.71 & 0.92 & 0.92 & 0.90 %
    & 0.83 & 0.96 & 0.84 & 0.98 & 0.97 & 0.97 %
    & 0.70 & 0.82 & 0.67 & 0.88 & 0.87 & 0.84 %
    & 0.64 & 0.85 & 0.65 & 0.85 & 0.88 & 0.83 %
\\

    \bottomrule
\end{tabular}
\end{adjustbox}

\end{table*}

We report the results of our experiment in Table~\ref{tab:scores_overall_combined}.
In this table, we include an overall assessment (column ``Overall''), which averages the results across the three datasets, and the results per dataset in their respective columns. 
We highlight the methods that achieved the best $F_1$ scores, and we will place special emphasis on this metric in our results discussion.

\textbf{Finding 1: The average accuracy of the methods varies considerably across the three different datasets}. 
In their default configuration setup, CPD methods reach an average $F_1$ of 0.84 in the \graal dataset, capturing 96\% of the \suspiciouschangepoints with 83\% precision. 
The \graal dataset is composed primarily of stationary series with a low coefficient of variance (RQ1), and often exhibits a single change point (RQ2). 
On the other hand, CPD methods performed significantly worse in the \mongo and \sap datasets, with an average $F_1$-score of 0.67 and 0.65, respectively, capturing $\approx$80\% of \suspiciouschangepoints but with much lower precision.

\textbf{Finding 2: Accuracy of CPD methods increases when optimally tuned.}
In the global average, the accuracy of CPD methods in the $F_1$-score increased 28\%, from 0.70 to 0.90.
The impact of fine-tuning more significantly affects the precision of the methods, increasing the average precision of CPD methods from 72\% to 92\%, essentially helping the CPD methods to better differentiate between the noise in the data and the annotated \suspiciouschangepoints.
Some methods, e.g., \texttt{KCPA} and \texttt{SEGNEIGH}, severely underperform in their default settings, suggesting that their configurations are not tuned to the characteristics of performance measurements. 
\texttt{KCPA}, for example, has a precision of only 3\%, essentially flagging most points in the performance series as potential change points; however, it reaches 86\% precision when properly tuned.

\textbf{Finding 3: \texttt{BINSEG}, reaches the best overall accuracy across all datasets.}
Across all evaluated CPD methods, \texttt{BINSEG} shows the best overall $F_1$-score, both on Default parameters and in the Oracle experimentation settings.
\texttt{BINSEG}is a simple method based on binary segmentation and maximum likelihood estimation~\cite{Scott1974}, but it achieves remarkable precision in our evaluation, with 94\% overall precision, while capturing 83\% of all suspicious change points.  
Across all our evaluations, \texttt{BINSEG} ranked either first or second, and it was the best method under default parameters, demonstrating robustness in its default settings.  
Our results indicate that, among the evaluated methods, \texttt{BINSEG} can be considered a safe first choice for performance regression detection.

\rev{R2C4}{
\textbf{Finding 4: \texttt{MONGO} provides strong out-of-the-box performance but benefits less from tuning.} \texttt{MONGO} achieves an overall $F_1$-score of $0.86$ with its default configuration, increasing to $0.91$ in the Oracle setting. On the MongoDB dataset itself, its $F_1$-score increases from $0.82$ to $0.88$. These results suggest that \texttt{MONGO} is comparatively robust without fine-tuning, but has a lower tuning ceiling than methods such as \texttt{BINSEG}, \texttt{BOCPD}, and \texttt{SEGNEIGH}.
This is consistent with the role of the method in its original deployment: E-Divisive Means was selected as a general-purpose, training-free CPD technique and integrated into a workflow in which detected changes are subsequently reviewed by engineers. 
Their production process is intentionally biased toward false positives, since missing a performance regression is considered more costly than investigating additional candidate changes.
Our evaluation instead measures precision and recall symmetrically through the $F_1$-score. This difference in evaluation objective helps explain why \texttt{MONGO} can be effective in its original operational setting without necessarily outperforming other CPD methods in our benchmark.
}

\textbf{Finding 5: \texttt{BOCPD} and \texttt{SEGNEIGH} showed comparable accuracy to \texttt{BINSEG} after parameter tuning.}
The \texttt{BOCPD} method showed average accuracy with its default parameters, but a significantly higher ceiling when parameter-tuned, achieving comparable accuracy to \texttt{BINSEG} across all datasets.
This is particularly acute in the \sap dataset.
In its default configuration, \texttt{BOCPD} exhibited very low precision (62\%), but when optimally tuned it reaches 97\% precision, accuracy comparable to the best methods. 
Another method with similar accuracy was \texttt{SEGNEIGH}, which showed abysmal out-of-the-box accuracy, with lows of 27\% precision in the \graal, but reached an overall $F_1$-score of 0.97 in Oracle mode. 
Both methods, \texttt{BOCPD} and \texttt{SEGNEIGH}, have low precision for performance data under their default parameters, due to their high sensitivity to data noise, but this can be tamed with hyperparameter tuning to focus on larger performance variations.

\textbf{Finding 6: CPD methods generally performed well on the \graal dataset.}
In the \graal dataset, \texttt{AMOC} performed the best with and without parameter tuning, but many others (\eg \texttt{BINSEG}, \texttt{MongoDB}, \texttt{PELT}) are very close.
\texttt{AMOC} only detects at most one change in the mean.
While this has the clear advantage of making it one of the fastest methods, it has the drawback of potentially hiding additional change points after the one with the largest magnitude of change.
On the \graal dataset, \texttt{AMOC} performed remarkably well because the considered time series seem to have very few change points, as also suggested by the human annotations.
However, in the more general case, it is probably advisable to give up some of the speed of \texttt{AMOC} and rely on any of the other methods that performed equally well but have the ability to detect more than one change.
For this dataset, it appears that hyperparameter tuning is not crucial, as most methods performed almost equally well with default parameters (with \texttt{KCPA} and \texttt{SEGNEIGH} being notable exceptions).

\section{Discussion and Implications}
\label{sec:discussion}
In this section, we discuss the key aspects of our results and some implications for practitioners and researchers.

\subsection{Characteristics of performance time series and CPD methods' accuracy.}\label{sec:char_vs_cpd}
Given the results on the accuracy of the change point detection methods presented in the previous section, we now investigate if there are specific methods that can be recommended based on the characteristics of the time series measurements.
This opens the possibility of prioritizing the methods based on the characteristics of the time series, which may be useful in practice to support a manual inspection and reduce the cost of the analysis.

\begin{figure}[htpb]
    \centering
    \includegraphics[width=.7\linewidth]{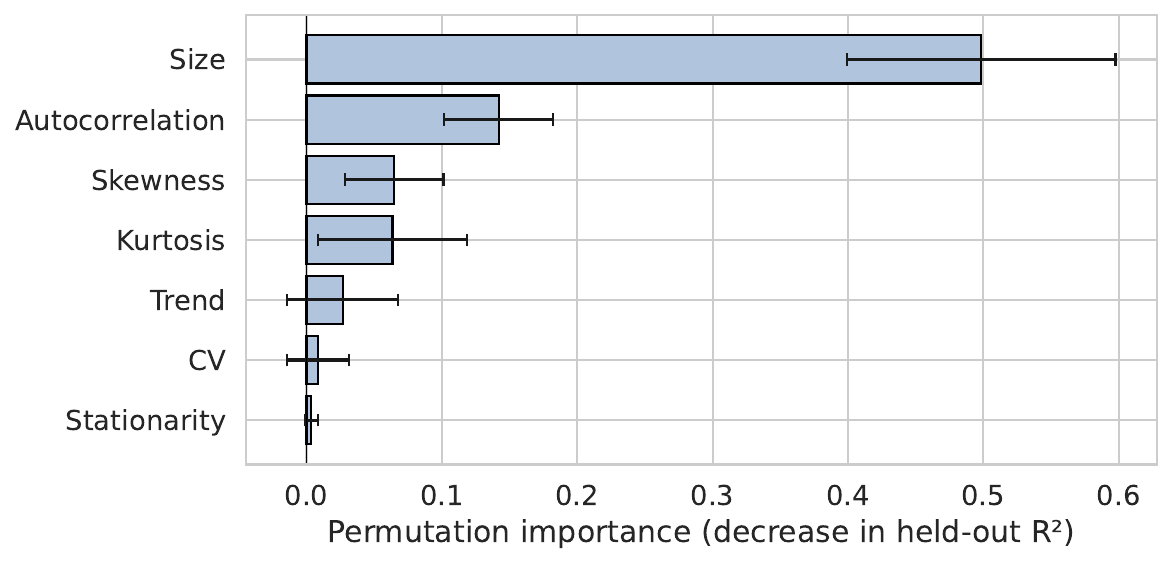}
    \caption{\rev{R2C7}{Random forest permutation feature importance of the characteristics of time-series measurements with respect to the average $F_1$-score achieved by the change point detection methods and configurations. Importance is measured as the decrease in held-out $R^2$ when a characteristic is permuted. Results are aggregated across 10 cross-validation folds with 30 permutation repetitions per fold. The model achieves $R^2 = 0.636 \pm 0.069$ and $\mathrm{RMSE} = 0.062 \pm 0.004$.}}
    \label{fig:cpd_and_char_random_forest_coefficients}
\end{figure}

\rev{R2C7}{We use a random forest regressor to investigate the relationship between time-series characteristics and the average $F_1$-score achieved by the CPD methods and configurations. We evaluate the model using 10-fold cross-validation and assess the contribution of each characteristic using permutation feature importance on the held-out folds. Permutation importance measures the decrease in predictive performance when the values of a feature are randomly permuted, thereby avoiding the known preference of impurity-based importance measures for features offering many potential split points. We repeat the permutation procedure 30 times per fold and report the mean importance and its variability across folds and repetitions. The model achieves an average $R^2$ of $0.636 \pm 0.069$ and an RMSE of $0.062 \pm 0.004$, indicating moderate predictive power.

Time-series size is by far the most influential characteristic in the model, with a permutation importance of $0.499 \pm 0.099$. This result is robust to the change in feature-importance method and supports the observation that CPD accuracy varies considerably with series length. Autocorrelation is the second most informative characteristic ($0.142 \pm 0.040$), followed by skewness ($0.065 \pm 0.036$) and kurtosis ($0.064 \pm 0.055$). In contrast, trend, coefficient of variation, and stationarity provide little additional predictive information once the other characteristics are considered.

One possible explanation for the association with series length is that longer time series have more candidate split points, which may increase the chances of over-segmentation and false positives. In addition, because most methods look for relatively stationary segments, slow trends and drifts in variance occurring over longer series may make change points harder to identify.
Autocorrelation also appears to be relevant to CPD accuracy. Strong temporal dependence can alter the apparent persistence of local changes and violates the independence assumptions underlying some detection procedures, potentially affecting their ability to distinguish structural changes from temporally correlated fluctuations.
Skewness and kurtosis also provide some predictive information, although their contribution is considerably smaller and less stable than that of series size and autocorrelation.
One possible explanation is that high kurtosis reflects a greater frequency of extreme values, which some methods may mistake for change points, while high skewness can occur alongside distributional asymmetries or trends that may make structural changes harder to distinguish.}

\paragraph{Early detection behavior of CPD methods}

When examining the detection patterns of the evaluated methods, we observe that many algorithms tend to report change points very early in the time series, often resulting in lower F1 scores. In contrast, detections occurring later in the series generally correspond to higher agreement with human annotations. This behavior likely reflects the statistical requirements of most CPD algorithms, which rely on observing sufficient post-change data to accumulate evidence of a distributional shift. Triggering too early may therefore lead to spurious detections before enough signal has emerged.

Another relevant factor is the frequency of cases where no change points are present. In these situations, some methods (notably \texttt{MONGO}, \texttt{CPNP}, \texttt{RFPOP}, and \texttt{WBS}) more closely match human annotations by correctly returning no detections, while other methods tend to over-segment the series. Taken together, these observations suggest that several CPD algorithms exhibit a bias toward early detection, whereas reliable detection typically requires a balance between accumulating enough post-change observations and avoiding overly aggressive triggering.

\subsection{Implications}

Our findings have implications for researchers investigating the application of change point detection (CPD) techniques on software performance data, and for practitioners responsible for performance regression analysis in industrial settings.

\paragraph{For researchers:}
The statistical properties of performance time series, namely low variance, weak trends, frequent extreme values, and limited autocorrelation (RQ1, Findings 1 and 2), indicate that CPD methods evaluated on datasets from other domains may not transfer reliably to software performance settings.
This suggests that future CPD research should emphasize robustness to heavy-tailed noise and avoid assumptions that rely strongly on trend structure or temporal dependence.

Although annotators exhibited high overall agreement when identifying potential performance changes (RQ2, Finding 1), the sharp reduction in shared change points as agreement thresholds increase (RQ2, Finding 2) reveals an inherent degree of subjectivity.
This implies that treating human annotations as a binary ground truth may be overly simplistic, motivating evaluation frameworks that account for uncertainty in expert judgments.

The observed negative correlation between annotation difficulty and inter-annotator agreement (RQ2, Finding 3) further suggests that difficulty can act as a proxy for time-series complexity.
Researchers may leverage this signal to stratify datasets and to study CPD accuracy under varying levels of ambiguity, rather than relying on aggregate scores alone.

Finally, the strong variation in CPD effectiveness across datasets and the substantial accuracy gains achieved through hyperparameter tuning (RQ3, Findings 1 and 2) indicate that no single CPD method is universally optimal. 
This opens opportunities for adaptive or meta-learning approaches that select and configure CPD methods based on measurable characteristics of the input time series (RQ1, Findings 1 and 2).

\paragraph{For practitioners:}
The combination of low overall noise and frequent extreme values in performance measurements (RQ1, Finding 1) implies that aggressive outlier removal may eliminate meaningful regression signals.
Practitioners should therefore exercise caution when preprocessing performance data, ensuring that potential change points are not inadvertently filtered out.

Despite high agreement among annotators in aggregate, many change points were identified by only a subset of experts (RQ2, Findings 1 and 2), which indicates that relying on a single reviewer or a single automated detector may be fragile.
Practitioners should instead aggregate multiple signals in their performance assurance processes to reduce the impact of individual bias or method-specific sensitivity.%
Moreover, since approximately one third of the time series were perceived as difficult or impossible to annotate (RQ2, Finding 3), CPD methods are best positioned as tools for prioritization rather than full automation. 
Automated detection can narrow the search space, while expert effort is reserved for ambiguous or high-risk cases.

Also, our results show that simple segmentation-based approaches such as BINSEG provide strong and stable out-of-the-box detection accuracy across datasets (RQ3, Finding 3), making them a pragmatic default when tuning resources or historical data are limited.
In contrast, recall-oriented methods (\ie CPD methods or configurations that prioritize detecting as many true change points as possible) may be preferable in contexts where the cost of missing a regression outweighs the cost of investigating false positives (RQ3, Finding 4). 
When resources allow for tuning, practitioners should consider leveraging the characteristics of their performance time series (RQ1, Findings 1 and 2) to guide method selection and configuration, as this can yield substantial improvements in detection accuracy (Section~\ref{sec:char_vs_cpd}). Additionally, while in this work we explored parameter tuning through an exhaustive grid search, practitioners may consider more efficient tuning strategies based on search heuristics like Bayesian optimization, Hyperband, or metaheuristic algorithms.

In practice, performance regression detection is typically performed in a continuous and incremental manner, where new performance measurements are analyzed as soon as they become available. Although our evaluation considers complete time series to enable controlled comparison across methods, these techniques can be applied in practice by re-running the detection periodically as new observations are collected. This incremental usage highlights the importance of timely detection, where practitioners aim to identify regressions as early as possible while still accumulating sufficient evidence to avoid spurious alerts.

\section{Threats to Validity}
\label{sec:threats}

In this section, we discuss potential threats to the validity of our study, organized according to four standard categories: construct, internal, external, and conclusion validity~\cite{WohlinRHORW24}.

\noindent
\textbf{Construct validity} concerns whether
the study measurements and operational definitions accurately capture the concepts under investigation.
A key construct threat arises from our definition of potential performance change, which reflects annotators' judgments rather than fully investigated or developer-confirmed regressions. As mitigation, multiple annotators labeled each series, and agreement was systematically quantified. Furthermore, the annotation tool's tutorial (see \secref{sec:annotation-tool}) introduced consistent rules for distinguishing noise, outliers, and change points.

Another construct threat concerns the use of a ±5-point margin of error for matching predicted and annotated change points (see \secref{sub:evaluation-metrics}).
Different tolerances could lead to different precision or recall values. We based this threshold on prior CPD evaluation work~\cite{Gerrit:2020:ChangePointDetection}, and we adopted it consistently across all methods and datasets to ensure comparability.

The characterization of time series using measures such as CoV, skewness, kurtosis, trend, stationarity, and autocorrelation represents another threat, as these characteristics may not fully capture nuances specific to performance measurements (\eg environment variability, CI-specific noise).
To mitigate this, we relied on characteristics widely used in time-series and CPD literature, which should enable cross-domain comparisons with established benchmarks such as Libra~\cite{Bauer:Libra:benchmark}.
\\

\noindent
\textbf{Internal validity} concerns factors that may affect the correctness of the study procedures or the causal relationships implied by findings.
A primary threat stems from the subjectivity of the manual annotation process. Although we recruited annotators with experience in performance engineering, identifying potential performance changes is inherently subjective. Differences in expertise, fatigue, or interpretation may influence labeling decisions.
To mitigate this threat, each time series was annotated by up to five independent annotators, and agreement was evaluated using OvR $F_1$-scores (see \figref{fig:ovr}), with generally high levels of consensus. We also enhanced the annotation tool to support task review, mandatory difficulty selection, and clearer UI prompts, as described in \secref{sec:annotation-tool}, aimed at reducing inconsistent labeling.

Another threat relates to the stratified sampling strategy used to select $1\,500$ time series for manual labeling. Because sampling was based on the number of change points detected by CPD methods, it may have introduced bias toward series that are more likely to exhibit detectable changes. As a mitigation, to ensure representation of both relatively stable and highly variable time series, we stratified across percentiles of unique change points, rather than absolute counts.

Parameters tuning for the CPD methods also introduces a potential internal threat. The Oracle configuration may overfit to the sampled time series, resulting in overly optimistic performance estimates. To control for this, we explicitly separate Default and Oracle configurations (see \tabref{tab:scores_overall_combined}), and interpret Oracle results as upper bounds rather than expected real-world performance.
\\

\noindent
\textbf{External validity}
External validity refers to the generalizability of findings beyond the datasets and settings studied.
Although our three industrial datasets (GraalVM, MongoDB, and SAP HANA) represent diverse types of software systems, they do not obviously encompass all performance-critical domains. For example, embedded systems, mobile environments, or systems with highly variable workload profiles may produce different types of performance measurements.
Also, our curation focused exclusively on time-based and throughput metrics (see \secref{sub:curation-process}), which limits generalizability to resource or event-based metrics such as memory consumption or I/O rate.

The exclusion of short time series ($<50$ data points) may limit applicability to organizations with sparse historical data. However, this decision aligns with established recommendations for CPD algorithms, many of which underperform on very short series. We made this restriction explicit when interpreting our results.
\\

\noindent
\textbf{Conclusion validity}
Conclusion validity addresses whether the study results follow logically from the collected evidence.
Our conclusions rely on a constructed ground truth based on human consensus, rather than definitive, human-validated performance regressions.
In this regard, high agreement among annotators does not guarantee that the labelled change points correspond to meaningful or user-impacting regressions in the underlying software systems.
Consequently, CPD methods that best approximate human perception (not necessarily operational relevance) rise to the top of our rankings. This distinction limits the strength of claims we can make about actual regression detection effectiveness. 

Another threat concerns the correlational analysis between time-series characteristics and CPD performance (see \figref{fig:cpd_and_char_random_forest_coefficients}).
This analysis is observational and correlational, conducted on a small set of datasets that share structural similarities (\eg CI-generated measurements). Therefore, it cannot establish causal or general rules about which characteristics truly determine CPD effectiveness. Confounding variables (e.g., dataset-specific test environments, measurement frequency, or instrumentation differences) cannot be fully ruled out.
As a consequence, we avoid causal claims and use the analyses primarily to generate hypotheses, not definitive conclusions.

Finally, another threat relates to the potential overinterpretation of tuning improvements. Oracle results are only to be interpreted as upper bounds of what can be achieved with expensive hyperparameter optimization, not as performance to be expected in practical deployments. As such, concluding that CPD methods \emph{can reach} certain $F_1$-scores may lead to choosing methods based on optimistic expectations. We mitigate this by clearly distinguishing Default and Oracle results throughout the paper.

\section{Conclusion}
\label{sec:conclusion}
In this study, we conducted a structured empirical evaluation of change point detection (CPD) methods in the context of software performance regression analysis. 
We curated three industrial datasets (GraalVM, MongoDB, SAP HANA) of software performance measurements as version-ordered time series through metric homogenization, exclusion of short series (< 50 observations), and removal of invalid measurements.
We first characterized the statistical properties of performance time series
and contrasted them with cross-domain benchmarks. Moreover, we established a reference ground truth by collecting structured annotations from performance experts, defining potential performance changes as data points warranting further investigation and quantifying inter-annotator agreement via one-vs-rest (OvR) $F_1$-scores under a $\pm5$-point matching tolerance. Finally, we evaluated 12 CPD algorithms spanning parametric, non-parametric, Bayesian, and segmentation-based approaches, measuring precision, recall, and $F_1$-score under both default configurations and exhaustively tuned (Oracle) parameter settings.

The results indicate that software performance time series exhibit comparatively low variance and weak trends, yet frequent heavy-tailed behavior and limited autocorrelation, distinguishing them from traditional CPD benchmark domains.
Human annotators showed high aggregate agreement (mean OvR $F_1$ $\approx$ 0.87), although stricter consensus thresholds revealed residual subjectivity.
CPD effectiveness varied substantially across datasets, with average default $F_1$-scores ranging from 0.65 to 0.84, and increased markedly under parameter tuning (average $F_1$ from 0.70 to 0.90), primarily through improved precision. 
Among the evaluated methods, binary segmentation (\texttt{BINSEG}) achieved the most stable and consistently high performance across datasets and configurations, while other methods exhibited strong sensitivity to hyperparameter settings.
Overall, the evidence suggests that CPD methods can approximate expert-identified performance changes under clearly specified assumptions, but their effectiveness depends critically on dataset characteristics and configuration choices.

Future work should further investigate several aspects of performance change detection in practice.
First, the quality and representativeness of the benchmarks used to evaluate application performance should be periodically reassessed. In many performance testing infrastructures, new benchmarks are continuously introduced while older ones remain in the suite without systematic reevaluation, potentially leading to outdated or redundant signals for regression detection.
Second, our results suggest that benchmark-specific tuning of change point detection methods could substantially improve accuracy, as the characteristics of performance time series may vary across different benchmarks and workloads. 
Finally, the evaluation of CPD methods in this study relied on the $F_1$-score, which implicitly assigns equal importance to false positives and false negatives. However, in the context of performance regression detection, the relative cost of these errors may differ depending on operational constraints. For instance, missing a regression may be more critical than investigating a false alarm, or vice versa. Exploring alternative evaluation metrics or cost-sensitive formulations that better reflect practical priorities represents an important direction for future research.

\begin{acks}
Diego is supported by the Natural Sciences and Engineering Research Council of Canada (NSERC) [funding reference number RGPIN-2023-05163].
\end{acks}

\bibliographystyle{ACM-Reference-Format}
\bibliography{bibliography}

\appendix

\end{document}